\documentclass[lettersize,journal]{IEEEtran}
\usepackage{amsmath,amsfonts}
\usepackage{amssymb}
\usepackage{subeqnarray}
\newtheorem{remark}{Remark}
\usepackage{cases}
\usepackage[ruled]{algorithm2e} 
\usepackage{array}
\usepackage[caption=false,font=normalsize,labelfont=sf,textfont=sf]{subfig}

\usepackage{stfloats}
\usepackage{url}

\usepackage{graphicx}
\usepackage{cite}
\begin{document}

\title{Minimum Error Entropy Rauch-Tung-Striebel Smoother (This work has been submitted to the IEEE for possible publication. Copyright may be transferred without notice, after which this version may no longer be accessible.)}

\author{Jiacheng He, Hongwei Wang, Gang Wang, Shan Zhong, Bei Peng

\thanks{This work was funded by the NNSFC with Nos. 51975107 and 62103083, Sichuan Science and Technology Major Project No.2019ZDZX0020, and Sichuan Science and Technology Program, No. 2022YFG0343. The first two authors contributed equally to this study. (Corresponding author: Bei Peng.)}
\thanks{J. He, S. Zhong, and B. Peng are with School of Mechanical and Electrical Engineering, University of Electronic Science and
Technology of China, Chengdu, 611731, PR China (e-mail: hejiacheng\_123@163.com; 2608589754@qq.com; beipeng@uestc.edu.cn).}
\thanks{H. Wang is with National Key Laboratory of Science and Technology on Communications, University of Electronic Science and Technology of China, Chengdu, 611731, PR China (e-mail: tianhangxinxiang@163.com).}
\thanks{G. Wang is with School of Information and Communication Engineering, University of Electronic Science and Technology of China, Chengdu, 611731, PR China (e-mail: wanggang\_hld@uestc.edu.cn).}
\thanks{Manuscript received April 19, 2021; revised August 16, 2021.}}

\markboth{Journal of \LaTeX\ Class Files,~Vol.~14, No.~8, August~2021}%
{Shell \MakeLowercase{\textit{et al.}}: A Sample Article Using IEEEtran.cls for IEEE Journals}


\maketitle

\begin{abstract}
Outliers and impulsive disturbances often cause heavy-tailed distributions in practical applications, and these will degrade the performance of Gaussian approximation smoothing algorithms. To improve the robustness of the Rauch-Tung-Striebel (RTS) smother against complicated non-Gaussian noises, a new RTS-smoother integrated with the minimum error entropy (MEE) criterion (MEE-RTS) is proposed for linear systems, which is also extended to the state estimation of nonlinear systems by utilizing the Taylor series linearization approach. The mean error behavior, the mean square error behavior, as well as the computational complexity of the MEE-RTS smoother are analyzed. According to simulation results, the proposed smoothers perform better than several robust solutions in terms of steady-state error.
\end{abstract}

\begin{IEEEkeywords}
Article submission, IEEE, IEEEtran, journal, \LaTeX, paper, template, typesetting.
\end{IEEEkeywords}

\section{Introduction}
\IEEEPARstart{S}{moothing} algorithms are important techniques for estimating the state of dynamic systems, which have been applied in various industrial applications such as target tracking \cite{9648188}, system identification \cite{LIU2021106995}, positioning \cite{CHIANG20112633}, and many others \cite{8571180,BALENZUELA2022110218,wang2016variational}. The Rauch-Tung-Striebel (RTS) smoother \cite{rauch1965maximum} is important for estimating the state of linear estimation with Gaussian noise. For nonlinear systems, some classic Gaussian smoothers are developed such as the extended RTS (ERTS) \cite{sarkka2013bayesian}, the unscented RTS (URTS) \cite{4484208}, the cubature Kalman \cite{ARASARATNAM20112245} smoothers and some iterative strategy-based smoothers \cite{bell1994iterated}. 

These aforementioned smoothers perform well with the presence of both the Gaussian process and measurement noises. However, a class of non-Gaussian measurement noise with heavy-tailed distributions is common in practical applications \cite{8540327,WANG2022108394,WANG20205058,ARAVKIN201763}. Non-Gaussian noise will dramatically decrease the performance of the smoothers that are initially devised for Gaussian noises. To address this issue, some strategies have been developed. Robust statistics, especially M-estimations \cite{WANG2019115,karlgaard2015nonlinear}, is one of the popular strategies to develop robust smoother for non-Gaussian noises. Another popular approach is to use the heavy-tailed distributions to model the non-Gaussian process and measurement noises, such as the Laplace \cite{5746504} distribution, Student's t  distribution \cite{7416166,7812899,9549725}, and generalized Gaussian scale mixture distribution \cite{8706940}. In addition, the robust smoothers based on ${H_\infty }$ \cite{5762390} and Gaussian mixture model have also been studied. In recent, the maximum correntropy criterion (MCC) \cite{4355325}, in information theoretic learning (ITL) \cite{5952087}, receives a great deal of attention, and some RTS-type smoothers are developed for nonlinear systems, such as maximum correntropy RTS (MC-RTS) \cite{wang2020maximum} smoother, maximum mixture correntropy nonlinear smoother (MMC-ORNS) \cite{LU2021108215}, and robust cubature Kalman smoother (RCKS) \cite{8540327}.

Although the MCC criterion is a good option for suppressing the effects of the heavy-tailed non-Gaussian noises \cite{8736038,9354208}, it may not perform as well for more complicated noise \cite{LU2021108215}, for example noises with multimodal distribution. In ITL, the minimum error entropy (MEE) criterion \cite{erdogmus2002error,HE2023109188}, as an important cost function, has been applied in adaptive filtering \cite{WANG2022108410,HE20221362,WANG2021107836,HE20223434}, Kalman filter (KF) \cite{9745084,CHEN2017,FENG2022108806}, and others \cite{silva2006error,santamaria2002entropy}. 

In addition, unscented KF \cite{dang2020robust} and cubature KF \cite{dang2021cubature} are also combined with the MEE criterion to enhance the estimation performance of the nonlinear systems with non-Gaussian noise. However, the MEE-based smoother has not been investigated. Moreover, compared with the KF-type algorithms, the smoothers based on the RTS framework has an additional step called backward pass, and this additional step gives the smoother an improved performance than the corresponding KF-type algorithms. Hence, a novel RTS smoother using the MEE criterion as a cost function is investigated, which is expected to obtain superior performance.

In this study,  a RTS-type smoother based the MEE criterion (MEE-RTS smoother) is proposed for the state estimation of the linear systems. Furthermore, an extended RTS smoother based on the MEE criterion is also developed for the nonlinear systems. Several theoretical analyses of the MEE-RTS smoother are presented, such as the mean error behavior, mean square error behavior, and computational complexity. Numerical simulations indicate the proposed smoothers outperform several existing robust smoothers in the mean-square deviation (MSD).

The remaining parts of this article are structured as follows. In Section \ref{background}, the problem formulation on the proposed smoothers are given. Two new smoothers on the basis of the MEE criterion are presented in Section \ref{mee_smoother}. A few theoretical analysis of the MEE-RTS smoother are presented in Section \ref{tanluysis}. Simulations, conclusions, and acknowledgments are presented in Section \ref{simulation}, Section \ref{conclusion}, and Section \ref{acknowledgments}.

\section{Problem formulation}\label{background}
A discrete-time linear systems, in this paper, is consider, and it is described using the following state-space model
\begin{align}\label{1qtrtxtj1fH}
\left\{ \begin{gathered}
  {{\boldsymbol{x}}_t} = {\boldsymbol{F}_t}{{\boldsymbol{x}}_{t - 1}} + {{\boldsymbol{q}}_{t}}, \hfill \\
  {{\boldsymbol{y}}_t} = {\boldsymbol{H}_t}{{\boldsymbol{x}}_t} + {{\boldsymbol{r}}_t}, \hfill \\ 
\end{gathered}  \right.
\end{align}
where  ${{\boldsymbol{x}}_t} \in {\mathbb{R}^{n \times 1}}$ represents the state of the linear systems at time $t$; ${{\boldsymbol{y}}_t} \in {\mathbb{R}^{m \times 1}}$ denotes the measurement of interest; ${\boldsymbol{F}_t}$ and ${\boldsymbol{H}_t}$ denote the state transition matrix and measurement matrix; ${{\boldsymbol{q}}_{t}}$ and ${{\boldsymbol{r}}_t}$ represent, respectively, the process and measurement noises. Nominally, both ${{\boldsymbol{q}}_{t}}$ and ${{\boldsymbol{r}}_t}$ follow zero-mean Gaussian distributions, and their covariance matrices are ${{\boldsymbol{Q}}_t}$ and ${{\boldsymbol{R}}_t}$, respectively. However, due to the impulse disturbance, outliers, or other factors, the distributions of ${{\boldsymbol{q}}_{t}}$ and ${{\boldsymbol{r}}_t}$ are generally no longer Gaussian, and reveal the heavy-tail properties. Such non-Gaussian distributions will degrade the performance of the existing RTS smoothers since they are initially devised under the Gaussian assumptions. To deal with the performance degradation, in this work, a robust smoother is developed to estimate the states ${{\boldsymbol{x}}_{1:T}} \triangleq \left\{ {{{\boldsymbol{x}}_1}, \cdots ,{{\boldsymbol{x}}_T}} \right\}$ utilizing the measurements ${{\boldsymbol{y}}_{1:T}} \triangleq \left\{ {{{\boldsymbol{y}}_1}, \cdots ,{{\boldsymbol{y}}_T}} \right\}$ with the presence of the heavy-tailed non-Gaussian noise. Specifically, we integrate the MEE criterion into the conventional RTS smoother framework to dampen the negative effect of the non-Gaussian noises.

\section{MEE based robust smoother}\label{mee_smoother}
\subsection{Minimum error entropy criterion}
The quadratic Renyi's entropy \cite{chen2019minimum}, also known as the error entropy, is the definition of 
\begin{align}\label{kekrvgolrhj1}
{H_2}\left( e \right) =  - \log {V_2}\left( e \right),
\end{align}
where error $e\triangleq X-Y$ denotes the error made between two random variables ${X}$ and ${Y}$, and the information potential ${V_2}\left( e \right)$ is given by
\begin{equation}\label{zkhxkhe1jrfpED}
\begin{split}
{V_2}\left( e \right) = \int {{p^2}\left( e \right)} de = {\text{E}}\left[ {p\left( e \right)} \right],
\end{split}
\end{equation}
where $p\left( e \right)$ represent the probability density function (PDF) on ${e}$, and ${{\text{E}}\left[  \cdot  \right]}$ is an expectation operator. The PDF ${p\left( e \right)}$ can be estimated via the Parzen window strategy employing the error samples:
\begin{equation}\label{xkhiejxxksgmg}
\begin{split}
\hat p\left( e \right) = \frac{1}{L}\sum\limits_{i = 1}^L G \left( {e - {e_i}} \right),
\end{split}
\end{equation}
where $G(e) = \left( {1/\sqrt {2\pi } \sigma } \right)\exp \left( { - {e^2}/2{\sigma ^2}} \right)$ is the Gaussian kernel function, and parameter ${\sigma }$ is the kernel bandwidth of $G\left(  \cdot  \right)$, and ${\left\{ {{e_i}} \right\}_{i = 1}^L}$ denote $L$ independent and identically distributed (i.i.d.) error samples. With the estimated PDF in \eqref{xkhiejxxksgmg}, one can obtain an estimate of ${{V_2}\left( e \right)}$ via \eqref{zkhxkhe1jrfpED} as
\begin{equation}\label{gsigeiejjiij}
{\hat V_2}\left( e \right) = \frac{1}{L}\sum\limits_{i = 1}^L {\hat p\left( e \right)}  = \frac{1}{{{L^2}}}\sum\limits_{i = 1}^L {\sum\limits_{j = 1}^L {G\left( {{e_i} - {e_j}} \right)} } ,
\end{equation}
and the detailed derivation process can be found in \cite{5952087}.

In ITL, minimizing the error entropy can be used as a criterion for optimization, i.e., the MEE criterion, which is given by
\begin{align}
\min H_2(e) = \max \log V_2(e).
\end{align} 
Due to the fact that $\max \log V_2(e)$ is equivalent to $\max V_2(e)$ and $V_2(e) $ can be estimated via (5) with samples, the MEE criterion can be further expressed as
\begin{align}
\max {V_2}(e) \approx \max \frac{1}{{{L^2}}}\sum\limits_{i = 1}^L {\sum\limits_{j = 1}^L G } \left( {{e_i} - {e_j}} \right).
\end{align}
The MEE criterion has unique advantages for handling complicated non-Gaussian noise \cite{chen2019minimum}, which motivates us to develop a new smoother based on the MEE criterion.

\subsection{The MEE based RTS smoother}
\subsubsection{Forward filtering procedure}
The forward filtering process is the foundation of the backward smoothing process, and its output $\left\{ {{{{\boldsymbol{\tilde x}}}_{t|t}},{{\boldsymbol{P}}_{t|t}}} \right\}$ are the input of the backward smoothing process. In the linear-regression-based KF solution, the measurement equation and filter update be recast as a regression problem \cite{karlgaard2015nonlinear}. Denote ${{\boldsymbol{x}}_t}$ the real state of the system, and hence the state prediction error can be expressed as ${{\boldsymbol{\delta }}_t} = {{\boldsymbol{\tilde x}}_{t|t - 1}} - {{\boldsymbol{x}}_t}$, where ${{\boldsymbol{\tilde x}}_{t|t - 1}}$ is the predicted state. With ${{{\boldsymbol{y}}_t}}$ and ${{{{\boldsymbol{\tilde x}}}_{t|t - 1}}}$, the regression problem with the form of
\begin{align}\label{gongshiXtjblh}
\left[ {\begin{array}{*{20}{c}}
  {{{\boldsymbol{y}}_t}} \\ 
  {{{{\boldsymbol{\tilde x}}}_{t|t - 1}}} 
\end{array}} \right] = \left[ {\begin{array}{*{20}{c}}
  {{{\boldsymbol{H}}_t}{{\boldsymbol{x}}_t}} \\ 
  {{{\boldsymbol{x}}_t}} 
\end{array}} \right] + \left[ {\begin{array}{*{20}{c}}
  {{{\boldsymbol{r}}_t}} \\ 
  {{{\boldsymbol{\delta }}_t}} 
\end{array}} \right].
\end{align}
Defining
\begin{align}
{{\boldsymbol{M}}_t} = \left[ {\begin{array}{*{20}{c}}
  {{\boldsymbol{R}}_t^{ - \tfrac{1}{2}}}&{\boldsymbol{0}} \\ 
  {\boldsymbol{0}}&{{\boldsymbol{P}}_{t|t - 1}^{ - \tfrac{1}{2}}} 
\end{array}} \right],
\end{align}
and multiplying ${{\boldsymbol{M}}_t}$ on both sides of \eqref{gongshiXtjblh} will transform the linear regression problem in \eqref{gongshiXtjblh} as
\begin{align}\label{8etjiatxTzD}
{{\boldsymbol{d}}_t} = {{\boldsymbol{Z}}_t}{{\boldsymbol{x}}_t} + {{\boldsymbol{e}}_t},
\end{align}
where 
\begin{align}\label{ytRpxttj1bla}
{{\boldsymbol{d}}_t} = \left[ {\begin{array}{*{20}{c}}
  {{\boldsymbol{R}}_t^{ - \tfrac{1}{2}}{{\boldsymbol{y}}_t}} \\ 
  {{\boldsymbol{P}}_{t|t - 1}^{ - \tfrac{1}{2}}{{{\boldsymbol{\tilde x}}}_{t|t - 1}}} 
\end{array}} \right],
\end{align}
\begin{align}\label{zTdzkrhrttsZk}
{{\boldsymbol{Z}}_t} = \left[ {\begin{array}{*{20}{c}}
  {{\boldsymbol{R}}_t^{ - \tfrac{1}{2}}{{\boldsymbol{H}}_t}} \\ 
  {{\boldsymbol{P}}_{t|t - 1}^{ - \tfrac{1}{2}}} 
\end{array}} \right],
\end{align}
and 
\begin{align}
{{\boldsymbol{e}}_t} = \left[ {\begin{array}{*{20}{c}}
  {{\boldsymbol{R}}_t^{ - \tfrac{1}{2}}}&{{{\boldsymbol{0}}_n}} \\ 
  {{{\boldsymbol{0}}_n}}&{{\boldsymbol{P}}_{t|t - 1}^{ - \tfrac{1}{2}}} 
\end{array}} \right]\left[ {\begin{array}{*{20}{c}}
  {{{\boldsymbol{r}}_t}} \\ 
  {{{\boldsymbol{\delta }}_t}} 
\end{array}} \right].
\end{align}
From \eqref{8etjiatxTzD}, the augmented error can be written as
\begin{align}\label{xtzTjTddeyTe}
{{\boldsymbol{e}}_t} = {{\boldsymbol{d}}_t} - {{\boldsymbol{Z}}_t}{{\boldsymbol{x}}_t},
\end{align}
and hence the optimal estimation of ${{\boldsymbol{\tilde x}}_{t|t}}$ can be gained by the MEE criterion, which is formulated as
\begin{align}\label{etfijitfeGsig}
{{{\boldsymbol{\tilde x}}}_{t|t}} &= \arg \mathop {\max }\limits_{{{\boldsymbol{x}}_t}} {J_{MEE}}\left( {{{\boldsymbol{x}}_t}} \right)\notag\\
 &= \arg \mathop {\max }\limits_{{{\boldsymbol{x}}_t}} \frac{1}{{{N^2}}}\sum\limits_{i = 1}^N {\sum\limits_{j = 1}^N {G\left( {{e_{t;i}} - {e_{t;j}}} \right)} } .
\end{align}
where ${e_{t;i}} = {d_{t;i}} - {{\boldsymbol{z}}_{t;i}}{{\boldsymbol{x}}_t}$, ${e_{t;i}}$ and ${d_{t;i}}$ are the $i$th element of ${{\boldsymbol{e}}_t}$ and ${{\boldsymbol{d}}_t}$, ${{\boldsymbol{z}}_{t;i}}$ represents the $i$th row of ${{\boldsymbol{Z}}_t}$. The variable $N$ is equal to $m + n$.

Set the gradient of the cost function in \eqref{etfijitfeGsig} on ${{{\boldsymbol{e}}_t}}$ will lead to
\begin{align}\label{zerodetetFai}
\frac{{\text{2}}}{{{N^2}{\sigma ^2}}}{\boldsymbol{Z}}_t^{\text{T}}{{\boldsymbol{\Psi }}_t}{{\boldsymbol{e}}_t} - \frac{{\text{2}}}{{{N^2}{\sigma ^2}}}{\boldsymbol{Z}}_t^{\text{T}}{{\boldsymbol{\Phi }}_t}{{\boldsymbol{e}}_t} = 0,
\end{align}
where the $\left( {i,j} \right)$th element of ${{\boldsymbol{\Psi }}_t}$ and ${{\boldsymbol{\Phi }}_t}$ are, respectively, given by
${\left[ {{{\boldsymbol{\Phi }}_t}} \right]_{ij}} = G\left( {{e_{t;i}} - {e_{t;j}}} \right)$ and 
\begin{align*}
{\left[ {{{\boldsymbol{\Psi }}_t}} \right]_{ij}} = \left\{ {\begin{array}{*{20}{l}}
  {\sum\limits_{j = 1}^N {G\left( {{e_{t;i}} - {e_{t;j}}} \right),} } \\ 
  {0,i \ne j,} 
\end{array}} \right.
\end{align*}
Substitute \eqref{xtzTjTddeyTe} into \eqref{zerodetetFai} and one can reach at
\begin{align}\label{DtOUtZtTj1zktZ}
{{\boldsymbol{x}}_t} = {\left[ {{\boldsymbol{Z}}_t^{\text{T}}{{\boldsymbol{\Omega }}_t}{{\boldsymbol{Z}}_t}} \right]^{ - 1}}{\boldsymbol{Z}}_t^{\text{T}}{{\boldsymbol{\Omega }}_t}{{\boldsymbol{d}}_t},
\end{align}
where ${{\boldsymbol{\Omega }}_t}$ is expressed as
\begin{align}\label{outfaizhzTjtfaizh}
{{\boldsymbol{\Omega }}_t} &= {\boldsymbol{\Psi }}_t^{\text{T}}{{\boldsymbol{\Psi }}_t} + {\boldsymbol{\Phi }}_t^{\text{T}}{{\boldsymbol{\Phi }}_t}  \notag\\
 &= \left[ {\begin{array}{*{20}{c}}
  {{{\boldsymbol{\Omega }}_{t;y}}}&{{{\boldsymbol{\Omega }}_{t;xy}}} \\ 
  {{{\boldsymbol{\Omega }}_{t;yx}}}&{{{\boldsymbol{\Omega }}_{t;x}}} 
\end{array}} \right].
\end{align}

Define ${\left[ {{\boldsymbol{Z}}_t^{\text{T}}{{\boldsymbol{\Omega }}_t}{{\boldsymbol{Z}}_t}} \right]^{ - 1}}{\boldsymbol{Z}}_t^{\text{T}}{{\boldsymbol{\Omega }}_t}{{\boldsymbol{d}}_t} = l\left( {{{\boldsymbol{x}}_t}} \right)$, and thus Eq. \eqref{DtOUtZtTj1zktZ} becomes a function on ${{\boldsymbol{x}}_t}$, so the estimated ${{{{\boldsymbol{\tilde x}}}_{t|t}}}$ can be obtained via the fixed-point iteration (FPI) method
\begin{align}
{\left( {{{{\boldsymbol{\tilde x}}}_{t|t}}} \right)_{k + 1}} = l\left[ {{{\left( {{{{\boldsymbol{\tilde x}}}_{t|t}}} \right)}_k}} \right] = {\left[ {{\boldsymbol{Z}}_t^{\text{T}}{{\left( {{{{\boldsymbol{\hat \Omega }}}_t}} \right)}_k}{{\boldsymbol{Z}}_t}} \right]^{ - 1}}{\boldsymbol{Z}}_t^{\text{T}}{\left( {{{{\boldsymbol{\hat \Omega }}}_t}} \right)_k}{{\boldsymbol{d}}_t},
\end{align}
where variables with superscript $ \wedge $ denote variables that are affected by the introduction of the FPI method, e.g. ${{{{\boldsymbol{\hat \Omega }}}_t}}$, where the subscript $k$ is the number of the FPI method.

Combining \eqref{ytRpxttj1bla}, \eqref{zTdzkrhrttsZk}, and \eqref{outfaizhzTjtfaizh} yields
\begin{align}\label{17ZtOtZtTDtOUtT}
\left\{ \begin{gathered}
  {\boldsymbol{Z}}_t^{\text{T}}{\left( {{{{\boldsymbol{\hat \Omega }}}_t}} \right)_k}{{\boldsymbol{Z}}_t} = {\boldsymbol{H}}_t^{\text{T}}{\left( {{\boldsymbol{\hat P}}_{t|t - 1}^y} \right)_k}{{\boldsymbol{H}}_t} + {\left( {{\boldsymbol{\hat P}}_{t|t - 1}^{yx}} \right)_k}{{\boldsymbol{H}}_t} \hfill \\
   + {\boldsymbol{H}}_t^{\text{T}}{\left( {{\boldsymbol{\hat P}}_{t|t - 1}^{xy}} \right)_k} + {\left( {{\boldsymbol{\hat P}}_{t|t - 1}^x} \right)_k}, \hfill \\
  {\boldsymbol{Z}}_t^{\text{T}}{\left( {{{{\boldsymbol{\hat \Omega }}}_t}} \right)_k}{{\boldsymbol{d}}_t} = \left[ {{\boldsymbol{H}}_t^{\text{T}}{{\left( {{\boldsymbol{\hat P}}_{t|t - 1}^y} \right)}_k} + {{\left( {{\boldsymbol{\hat P}}_{t|t - 1}^{yx}} \right)}_k}} \right]{{\boldsymbol{y}}_t} \hfill \\
   + \left[ {{\boldsymbol{H}}_t^{\text{T}}{{\left( {{\boldsymbol{\hat P}}_{t|t - 1}^{xy}} \right)}_k} + {{\left( {{\boldsymbol{\hat P}}_{t|t - 1}^x} \right)}_k}} \right]{{{\boldsymbol{\tilde x}}}_{t|t - 1}}. \hfill \\ 
\end{gathered}  \right.
\end{align}
with
\begin{align}\label{PPPPxy}
\left\{ \begin{gathered}
  {\left( {{\boldsymbol{\hat P}}_{t|t - 1}^x} \right)_k} = {\left[ {{\boldsymbol{P}}_{t|t - 1}^{ - \tfrac{1}{2}}} \right]^{\text{T}}}{\left( {{{{\boldsymbol{\hat \Omega }}}_{t;x}}} \right)_k}{\boldsymbol{P}}_{t|t - 1}^{ - \tfrac{1}{2}}, \hfill \\
  {\left( {{\boldsymbol{\hat P}}_{t|t - 1}^{xy}} \right)_k} = {\left[ {{\boldsymbol{R}}_t^{ - \tfrac{1}{2}}} \right]^{\text{T}}}{\left( {{{{\boldsymbol{\hat \Omega }}}_{t;xy}}} \right)_k}{\boldsymbol{P}}_{t|t - 1}^{ - \tfrac{1}{2}}, \hfill \\
  {\left( {{\boldsymbol{\hat P}}_{t|t - 1}^{yx}} \right)_k} = {\left[ {{\boldsymbol{P}}_{t|t - 1}^{ - \tfrac{1}{2}}} \right]^{\text{T}}}{\left( {{{{\boldsymbol{\hat \Omega }}}_{t;yx}}} \right)_k}{\boldsymbol{R}}_t^{ - \tfrac{1}{2}}, \hfill \\
  {\left( {{\boldsymbol{\hat P}}_{t|t - 1}^y} \right)_k} = {\left[ {{\boldsymbol{R}}_t^{ - \tfrac{1}{2}}} \right]^{\text{T}}}{\left( {{{{\boldsymbol{\hat \Omega }}}_{t;y}}} \right)_k}{\boldsymbol{R}}_t^{ - \tfrac{1}{2}}. \hfill \\ 
\end{gathered}  \right.
\end{align}
Substituting \eqref{17ZtOtZtTDtOUtT} into \eqref{DtOUtZtTj1zktZ} yields the update of ${\left( {{{{\boldsymbol{\tilde x}}}_{t|t}}} \right)_{k + 1}}$ as
\begin{align}\label{XtKtYtjHx}
{\left( {{{{\boldsymbol{\tilde x}}}_{t|t}}} \right)_{k + 1}} = {{{\boldsymbol{\tilde x}}}_{t|t - 1}} + {\left( {{{{\boldsymbol{\hat K}}}_t}} \right)_k}\left( {{{\boldsymbol{y}}_t} - {{\boldsymbol{H}}_t}{{{\boldsymbol{\tilde x}}}_{t|t - 1}}} \right),
\end{align}
where the Kalman gain is
\begin{align}\label{KgainKj1yjZoTzK}
{\left( {{{{\boldsymbol{\hat K}}}_t}} \right)_k} = {\left[ {{\boldsymbol{Z}}_t^{\text{T}}{{\left( {{{{\boldsymbol{\hat \Omega }}}_t}} \right)}_k}{{\boldsymbol{Z}}_t}} \right]^{ - 1}}\left[ {{{\left( {{\boldsymbol{\hat P}}_{t|t - 1}^{yx}} \right)}_k} + {\boldsymbol{H}}_t^{\text{T}}{{\left( {{\boldsymbol{\hat P}}_{t|t - 1}^y} \right)}_k}} \right].
\end{align}
When the condition ${{\left\| {{{\left( {{{{\boldsymbol{\tilde x}}}_{t|t}}} \right)}_{k + 1}} - {{\left( {{{{\boldsymbol{\tilde x}}}_{t|t}}} \right)}_k}} \right\|} \mathord{\left/
 {\vphantom {{\left\| {{{\left( {{{{\boldsymbol{\tilde x}}}_{t|t}}} \right)}_{k + 1}} - {{\left( {{{{\boldsymbol{\tilde x}}}_{t|t}}} \right)}_k}} \right\|} {\left\| {{{\left( {{{{\boldsymbol{\tilde x}}}_{t|t}}} \right)}_k}} \right\|}}} \right.
 \kern-\nulldelimiterspace} {\left\| {{{\left( {{{{\boldsymbol{\tilde x}}}_{t|t}}} \right)}_k}} \right\|}} \leqslant \tau $ is satisfied, the FPI algorithm stops, and ${{\boldsymbol{\hat K}}_t} = {\left( {{{{\boldsymbol{\hat K}}}_t}} \right)_k}$.
The covariance matrix ${{\boldsymbol{P}}_{t|t}}$ can be updated in the following way
\begin{align}\label{pTtIrKtTjTjIdanwei}
{{\boldsymbol{P}}_{t|t}} = \left( {{\boldsymbol{I}} - {{{\boldsymbol{\hat K}}}_t}{{\boldsymbol{H}}_t}} \right){{\boldsymbol{P}}_{t|t - 1}}{\left( {{\boldsymbol{I}} - {{{\boldsymbol{\hat K}}}_t}{{\boldsymbol{H}}_t}} \right)^{\text{T}}} + {{\boldsymbol{\hat K}}_t}{{\boldsymbol{R}}_t}{\boldsymbol{\hat K}}_t^{\text{T}}.
\end{align}

\subsubsection{Backward smoothing procedure}
The backward smoothing procedure is performed after the forward filtering procedure. The objective of the backward smoothing procedure is the smoothing state ${{\boldsymbol{\tilde x}}_{t|T}}\left( {t = T - 1,T - 2, \cdots ,1} \right)$ and covariance
matrices ${{\boldsymbol{P}}_{t|T}}\left( {t = T - 1,T - 2, \cdots ,1} \right)$.

Similar to the forward pass, the backward smoothing problem can be reformulated as a linear regression problem based on the smoothing state ${{{\boldsymbol{\tilde x}}}_{t + 1|T}}$ and the filtering state ${{{\boldsymbol{\tilde x}}}_{t|t}}$. The regression problem takes the form
\begin{align}\label{routxtFtjgat}
\left[ {\begin{array}{*{20}{c}}
  {{{{\boldsymbol{\tilde x}}}_{t + 1|T}}} \\ 
  {{{{\boldsymbol{\tilde x}}}_{t|t}}} 
\end{array}} \right] = \left[ {\begin{array}{*{20}{c}}
  {{{\boldsymbol{F}}_t}{{\boldsymbol{x}}_t}} \\ 
  {{{\boldsymbol{x}}_t}} 
\end{array}} \right] + \left[ {\begin{array}{*{20}{c}}
  {{{\boldsymbol{\rho }}_t}} \\ 
  {{{\boldsymbol{\gamma }}_t}} 
\end{array}} \right],
\end{align}
where ${{\boldsymbol{\gamma }}_t} = {{{\boldsymbol{\tilde x}}}_{t|t}} - {{\boldsymbol{x}}_t}$ denotes the estimation error of ${{{\boldsymbol{\tilde x}}}_{t|t}}$, and ${{\boldsymbol{\rho }}_t} = {{{\boldsymbol{\tilde x}}}_{t + 1|T}} - {{\boldsymbol{F}}_t}{{\boldsymbol{x}}_t}$ represents the prediction error of the state.

Define
\begin{align}
{{\boldsymbol{Y}}_t} = \left[ {\begin{array}{*{20}{c}}
  {{\boldsymbol{Q}}_t^{ - \tfrac{1}{2}}}&{\boldsymbol{0}} \\ 
  {\boldsymbol{0}}&{{\boldsymbol{P}}_{t|t}^{ - \tfrac{1}{2}}} 
\end{array}} \right].
\end{align}
Both sides of \eqref{routxtFtjgat} are simultaneously multiplied together by ${{\boldsymbol{Y}}_t}$, and we can get
\begin{align}\label{thetadeyuwtxtjia}
{{\boldsymbol{\Theta }}_t} = {{\boldsymbol{W}}_t}{{\boldsymbol{x}}_t} + {{\boldsymbol{\varepsilon }}_t}
\end{align}
with 
\begin{align}\label{14xtxiT-12}
{{\boldsymbol{\Theta }}_t} = \left[ {\begin{array}{*{20}{c}}
  {{\boldsymbol{Q}}_t^{ - \tfrac{1}{2}}{{{\boldsymbol{\tilde x}}}_{t + 1|T}}} \\ 
  {{\boldsymbol{P}}_{t|t}^{ - \tfrac{1}{2}}{{{\boldsymbol{\tilde x}}}_{t|t}}} 
\end{array}} \right],
\end{align}
\begin{align}\label{FP12fqt}
{{\boldsymbol{W}}_t} = \left[ {\begin{array}{*{20}{c}}
  {{\boldsymbol{Q}}_t^{ - \tfrac{1}{2}}{{\boldsymbol{F}}_t}} \\ 
  {{\boldsymbol{P}}_{t|t}^{ - \tfrac{1}{2}}} 
\end{array}} \right],
\end{align}
and
\begin{align}
{{\boldsymbol{\varepsilon }}_t} = \left[ {\begin{array}{*{20}{c}}
  {{\boldsymbol{Q}}_t^{ - \tfrac{1}{2}}}&{\boldsymbol{0}} \\ 
  {\boldsymbol{0}}&{{\boldsymbol{P}}_{t|t}^{ - \tfrac{1}{2}}} 
\end{array}} \right]\left[ {\begin{array}{*{20}{c}}
  {{{\boldsymbol{\rho }}_t}} \\ 
  {{{\boldsymbol{\gamma }}_t}} 
\end{array}} \right].
\end{align}
From \eqref{thetadeyuwtxtjia}, the augmented error can be written as
\begin{align}\label{thetajwtxt}
{{\boldsymbol{\varepsilon }}_t} = {{\boldsymbol{\Theta }}_t} - {{\boldsymbol{W}}_t}{{\boldsymbol{x}}_t}.
\end{align}

Based on the MEE criterion, the optimal estimation of ${{{\boldsymbol{x}}_t}}$ is obtained via
\begin{align}\label{etijetjbbjG}
\begin{gathered}
  {{{\boldsymbol{\tilde x}}}_{t|T}} = \arg \mathop {\max }\limits_{{{\boldsymbol{x}}_t}} {J_{MEE}}\left( {{{\boldsymbol{x}}_t}} \right) \hfill \\
   = \arg \mathop {\max }\limits_{{{\boldsymbol{x}}_t}} \frac{1}{{{L^2}}}\sum\limits_{i = 1}^L {\sum\limits_{j = 1}^L {G\left( {{\varepsilon _{t;i}} - {\varepsilon _{t;j}}} \right)} } , \hfill \\ 
\end{gathered} 
\end{align}
where ${\varepsilon _{t;i}} = {\theta _{t;i}} - {{\boldsymbol{w}}_{t;i}}{{\boldsymbol{x}}_t}$, ${\varepsilon _{t;i}}$ and ${\theta _{t;i}}$ are the $i$th element of ${{{\boldsymbol{\varepsilon }}_t}}$ and ${{{\boldsymbol{\Theta }}_t}}$, ${{\boldsymbol{w}}_{t;i}}$ represents the $i$th row of ${{\boldsymbol{W}}_t}$, and $L = 2n$ is the dimension of ${{{\boldsymbol{\varepsilon }}_t}}$.
The gradient of the cost function in \eqref{etijetjbbjG} can be expressed as:
\begin{align}\label{16tijbgtTWTLI}
\frac{{\partial {J_{MEE}}\left( {{{\boldsymbol{x}}_t}} \right)}}{{\partial {{\boldsymbol{x}}_t}}} = \frac{{\text{2}}}{{{L^2}{\sigma ^2}}}\sum\limits_{i = 1}^L {\sum\limits_{j = 1}^L {\left[ {{\boldsymbol{w}}_{t;j}^{\text{T}}{{G}}\left( {{\varepsilon _{t;j}} - {\varepsilon _{t;i}}} \right){\varepsilon _{t;j}}} \right]} } \notag\\
 - \frac{{\text{2}}}{{{L^2}{\sigma ^2}}}\sum\limits_{i = 1}^L {\sum\limits_{j = 1}^L {\left[ {{\boldsymbol{w}}_{t;j}^{\text{T}}{{G}}\left( {{\varepsilon _{t;j}} - {\varepsilon _{t;i}}} \right){\varepsilon _{t;i}}} \right].} } 
\end{align}
To further simplify, Eq. \eqref{16tijbgtTWTLI} can be further written as
\begin{align}\label{17edtbesL}
\frac{{\partial {J_{MEE}}\left( {{{\boldsymbol{x}}_t}} \right)}}{{\partial {{\boldsymbol{x}}_t}}}{\text{ = }}\frac{{\text{2}}}{{{L^2}{\sigma ^2}}}{\boldsymbol{W}}_t^{\text{T}}{{\boldsymbol{\Gamma }}_t}{{\boldsymbol{\varepsilon }}_t} - \frac{{\text{2}}}{{{L^2}{\sigma ^2}}}{\boldsymbol{W}}_t^{\text{T}}{{\boldsymbol{\Lambda }}_t}{{\boldsymbol{\varepsilon }}_t}
\end{align}
with
\begin{align}\label{FFigmatrix}
{\left[ {{{\boldsymbol{\Gamma }}_t}} \right]_{ij}} = \left\{ {\begin{array}{*{20}{l}}
  {\sum\limits_{j = 1}^L {{{G}}\left( {{\varepsilon _{t;i}} - {\varepsilon _{t;j}}} \right),i = j,} } \\ 
  {0,{\text{                       }}i \ne j,} 
\end{array}} \right.
\end{align}
and
\begin{align}\label{Getigjetij}
{\left[ {{{\boldsymbol{\Lambda }}_t}} \right]_{ij}} = {{G}}\left( {{\varepsilon _{t;i}} - {\varepsilon _{t;j}}} \right),
\end{align}
where ${\left[ {{{\boldsymbol{\Gamma }}_t}} \right]_{ij}}$ and ${\left[ {{{\boldsymbol{\Lambda }}_t}} \right]_{ij}}$ are the elements of the matrices ${{\boldsymbol{\Gamma }}_t}$ and ${{\boldsymbol{\Lambda }}_t}$, respectively.

To calculate ${{\boldsymbol{\tilde x}}_{t|T}}$ using \eqref{etijetjbbjG}, Eq. \eqref{17edtbesL} is set to zero, and a numerically stable solution \cite{wang2021numerically} can be obtained
\begin{align}\label{xtbwbfgdfd}
\left\{ {\begin{array}{*{20}{c}}
  {{{\boldsymbol{\Gamma }}_t}{{\boldsymbol{\varepsilon }}_t}{\text{ = }}{{\boldsymbol{\Gamma }}_t}\left[ {{{\boldsymbol{\Theta }}_t} - {{\boldsymbol{W}}_t}{{\boldsymbol{x}}_t}} \right]{\text{ = 0,}}} \\ 
  {{{\boldsymbol{\Lambda }}_t}{{\boldsymbol{\varepsilon }}_t}{\text{ = }}{{\boldsymbol{\Lambda }}_t}\left[ {{{\boldsymbol{\Theta }}_t} - {{\boldsymbol{W}}_t}{{\boldsymbol{x}}_t}} \right]{\text{ = 0.}}} 
\end{array}} \right.
\end{align}
It is obvious that \eqref{xtbwbfgdfd} can be further written as
\begin{align}\label{wbgsfsdtDBT}
\left[ {\begin{array}{*{20}{c}}
  {{{\boldsymbol{\Gamma }}_t}} \\ 
  {{{\boldsymbol{\Lambda }}_t}} 
\end{array}} \right]{{\boldsymbol{\Theta }}_t}{\text{ = }}\left[ {\begin{array}{*{20}{c}}
  {{{\boldsymbol{\Gamma }}_t}} \\ 
  {{{\boldsymbol{\Lambda }}_t}} 
\end{array}} \right]{{\boldsymbol{W}}_t}{{\boldsymbol{x}}_t}.
\end{align}
Multiply both sides of \eqref{wbgsfsdtDBT} using ${\boldsymbol{W}}_t^{\text{T}}\left[ {\begin{array}{*{20}{c}}
  {{\boldsymbol{\Gamma }}_t^{\text{T}}}&{{\boldsymbol{\Lambda }}_t^{\text{T}}} 
\end{array}} \right]$, and we can get
\begin{align}
{\boldsymbol{W}}_t^{\text{T}}\left( {{\boldsymbol{\Gamma }}_t^{\text{T}}{{\boldsymbol{\Gamma }}_t} + {\boldsymbol{\Lambda }}_t^{\text{T}}{{\boldsymbol{\Lambda }}_t}} \right){{\boldsymbol{\Theta }}_t} = {\boldsymbol{W}}_t^{\text{T}}\left( {{\boldsymbol{\Gamma }}_t^{\text{T}}{{\boldsymbol{\Gamma }}_t} + {\boldsymbol{\Lambda }}_t^{\text{T}}{{\boldsymbol{\Lambda }}_t}} \right){{\boldsymbol{W}}_t}{{\boldsymbol{x}}_t}. 
\end{align}
Further, the state of the system ${{\boldsymbol{x}}_t}$ can be obtained by
\begin{align}\label{theasanttTwSAN}
{{\boldsymbol{x}}_t} = {\left( {{\boldsymbol{W}}_t^{\text{T}}{{\boldsymbol{\Xi }}_t}{{\boldsymbol{W}}_t}} \right)^{ - 1}}{\boldsymbol{W}}_t^{\text{T}}{{\boldsymbol{\Xi }}_t}{{\boldsymbol{\Theta }}_t}
\end{align}
with
\begin{align}\label{ffAATT}
{{\boldsymbol{\Xi }}_t} = {\boldsymbol{\Gamma }}_t^{\text{T}}{{\boldsymbol{\Gamma }}_t} + {\boldsymbol{\Lambda }}_t^{\text{T}}{{\boldsymbol{\Lambda }}_t}.
\end{align}

Since ${\left( {{\boldsymbol{W}}_t^{\text{T}}{{\boldsymbol{\Xi }}_t}{{\boldsymbol{W}}_t}} \right)^{ - 1}}{\boldsymbol{W}}_t^{\text{T}}{{\boldsymbol{\Xi }}_t}{{\boldsymbol{\Theta }}_t}$ is a function of ${{\boldsymbol{x}}_t}$, Eq. \eqref{theasanttTwSAN} can be further written as 
\begin{align}
{{\boldsymbol{x}}_t} = {\boldsymbol{g}}\left( {{{\boldsymbol{x}}_t}} \right){\text{ = }}{\left( {{\boldsymbol{W}}_t^{\text{T}}{{\boldsymbol{\Xi }}_t}{{\boldsymbol{W}}_t}} \right)^{ - 1}}{\boldsymbol{W}}_t^{\text{T}}{{\boldsymbol{\Xi }}_t}{{\boldsymbol{\Theta }}_t}.
\end{align}

To obtain the numerical solution of ${{\boldsymbol{x}}_t}$, two common methods are often used. One solution in \cite{wang2020maximum,WANG20178659} is to use an approximate replacement method (ARM) that replaces ${{\boldsymbol{x}}_t}$ contained in ${\varepsilon _{t;i}}$ by ${{\boldsymbol{\tilde x}}_{t|t}}$. The other solution is to use the FPI method to calculate the smoothing estimation ${{{\boldsymbol{\tilde x}}}_{t|T}}$, and one can obtain
\begin{align}
{\left( {{{{\boldsymbol{\tilde x}}}_{t|T}}} \right)_k}{\text{ = }}{\left[ {{\boldsymbol{W}}_t^{\text{T}}{{\left( {{{{\boldsymbol{\hat \Xi }}}_t}} \right)}_k}{{\boldsymbol{W}}_t}} \right]^{ - 1}}{\boldsymbol{W}}_t^{\text{T}}{\left( {{{{\boldsymbol{\hat \Xi }}}_t}} \right)_k}{{\boldsymbol{\Theta }}_t},
\end{align}
where $k$ denotes the number of the FPI method, and the initial value of the FPI is ${\left( {{{{\boldsymbol{\tilde x}}}_{t|T}}} \right)_{k = 0}} = {{{\boldsymbol{\tilde x}}}_{t|t}}$. When only one FPI loop is performed, the FPI turns into the ARM in \cite{wang2020maximum}, which indicates that the ARM is a special case of the FPI method.

Whether FPI or ARM is used, ${{\boldsymbol{x}}_t}$ in (14) will be replaced by ${{{\boldsymbol{\tilde x}}}_{t|t}}$. The superscript $ \wedge $ is used to indicate the variable that has changed as a result of this substitution such as ${\hat \varepsilon _{t;i}}$, ${{\boldsymbol{\hat \Xi }}_t}$ etc.

To simplify the following derivation, ${\left( {{{{\boldsymbol{\hat \Xi }}}_t}} \right)_k}$ is written in the form of a block matrix as follows
\begin{align}\label{ssssttt11122122}
{\left( {{{{\boldsymbol{\hat \Xi }}}_t}} \right)_k}{\text{ = }}\left[ {\begin{array}{*{20}{c}}
  {{{\left( {{{{\boldsymbol{\hat \Xi }}}_{t;{x_1}}}} \right)}_k}}&{{{\left( {{{{\boldsymbol{\hat \Xi }}}_{t;{x_1}{x_2}}}} \right)}_k}} \\ 
  {{{\left( {{{{\boldsymbol{\hat \Xi }}}_{t;{x_2}{x_1}}}} \right)}_k}}&{{{\left( {{{{\boldsymbol{\hat \Xi }}}_{t;{x_2}}}} \right)}_k}} 
\end{array}} \right],
\end{align}
where ${{\boldsymbol{\hat \Xi }}_{t;{x_1}}} \in {\mathbb{R}^{n \times n}}$, ${{\boldsymbol{\hat \Xi }}_{t;{x_1}{x_2}}} \in {\mathbb{R}^{n \times n}}$, ${{\boldsymbol{\hat \Xi }}_{t;{x_2}{x_1}}} \in {\mathbb{R}^{n \times n}}$, and ${{\boldsymbol{\hat \Xi }}_{t;{x_2}}} \in {\mathbb{R}^{n \times n}}$.

Substitute \eqref{FP12fqt} and \eqref{ssssttt11122122} into ${\boldsymbol{W}}_t^{\text{T}}{({{{\boldsymbol{\hat \Xi }}}_t})_k}{{\boldsymbol{W}}_t}$ yields
\begin{align}\label{30wwswtT}
\begin{gathered}
  {\boldsymbol{W}}_t^{\text{T}}{\left( {{{{\boldsymbol{\hat \Xi }}}_t}} \right)_k}{{\boldsymbol{\Theta }}_t} = \left[ {{\boldsymbol{F}}_t^{\text{T}}{{\left( {{\boldsymbol{\hat P}}_{t|t - 1}^{b;{x_1}}} \right)}_k} + {{\left( {{\boldsymbol{\hat P}}_{t|t - 1}^{b;{x_2}{x_1}}} \right)}_k}} \right] \hfill \\
   \times {{{\boldsymbol{\tilde x}}}_{t + 1|T}} + \left[ {{\boldsymbol{F}}_t^{\text{T}}{{\left( {{\boldsymbol{\hat P}}_{t|t - 1}^{b;{x_1}{x_2}}} \right)}_k} + {{\left( {{\boldsymbol{\hat P}}_{t|t - 1}^{b;{x_2}}} \right)}_k}} \right]{{{\boldsymbol{\tilde x}}}_{t|t}}. \hfill \\ 
\end{gathered} 
\end{align}
with
\begin{align}\label{27PPPPx12}
\left\{ \begin{gathered}
  {\left( {{\boldsymbol{\hat P}}_{t|t - 1}^{b;{x_1}}} \right)_k} = {\left( {{\boldsymbol{Q}}_t^{ - \tfrac{1}{2}}} \right)^{\text{T}}}{\left( {{{{\boldsymbol{\hat \Xi }}}_{t;{x_1}}}} \right)_k}{\boldsymbol{Q}}_t^{ - \tfrac{1}{2}}, \hfill \\
  {\left( {{\boldsymbol{\hat P}}_{t|t - 1}^{b;{x_2}}} \right)_k} = {\left( {{\boldsymbol{P}}_{t|t}^{ - \tfrac{1}{2}}} \right)^{\text{T}}}{\left( {{{{\boldsymbol{\hat \Xi }}}_{t;{x_2}}}} \right)_k}{\boldsymbol{P}}_{t|t}^{ - \tfrac{1}{2}}, \hfill \\
  {\left( {{\boldsymbol{\hat P}}_{t|t - 1}^{b;{x_1}{x_2}}} \right)_k} = {\left( {{\boldsymbol{Q}}_t^{ - \tfrac{1}{2}}} \right)^{\text{T}}}{\left( {{{{\boldsymbol{\hat \Xi }}}_{t;{x_1}{x_2}}}} \right)_k}{\boldsymbol{P}}_{t|t}^{ - \tfrac{1}{2}}, \hfill \\
  {\left( {{\boldsymbol{\hat P}}_{t|t - 1}^{b;{x_2}{x_1}}} \right)_k} = {\left( {{\boldsymbol{P}}_{t|t}^{ - \tfrac{1}{2}}} \right)^{\text{T}}}{\left( {{{{\boldsymbol{\hat \Xi }}}_{t;{x_2}{x_1}}}} \right)_k}{\boldsymbol{Q}}_t^{ - \tfrac{1}{2}}. \hfill \\ 
\end{gathered}  \right.
\end{align}
Similar to the process of obtaining \eqref{30wwswtT}, Eq. \eqref{DbtStTWbt} can be obtained by substituting \eqref{14xtxiT-12}, \eqref{FP12fqt}, and \eqref{ssssttt11122122} into ${\boldsymbol{W}}_t^{\text{T}}{({{{\boldsymbol{\hat \Xi }}}_t})_k}{{\boldsymbol{\Theta }}_t}$
\begin{align}\label{DbtStTWbt}
\begin{gathered}
  {\boldsymbol{W}}_t^{\text{T}}{\left( {{{{\boldsymbol{\hat \Xi }}}_t}} \right)_k}{{\boldsymbol{\Theta }}_t} = \left[ {{\boldsymbol{F}}_t^{\text{T}}{{\left( {{\boldsymbol{\hat P}}_{t|t - 1}^{b;{x_1}}} \right)}_k} + {{\left( {{\boldsymbol{\hat P}}_{t|t - 1}^{b;{x_2}{x_1}}} \right)}_k}} \right] \hfill \\
   \times {{{\boldsymbol{\tilde x}}}_{t + 1|T}} + \left[ {{\boldsymbol{F}}_t^{\text{T}}{{\left( {{\boldsymbol{\hat P}}_{t|t - 1}^{b;{x_1}{x_2}}} \right)}_k} + {{\left( {{\boldsymbol{\hat P}}_{t|t - 1}^{b;{x_2}}} \right)}_k}} \right]{{{\boldsymbol{\tilde x}}}_{t|t}}. \hfill \\ 
\end{gathered}  
\end{align}
The matrix inversion lemma \cite{alexander2012adaptive} is utilized, and the following definitions
\begin{align}
\left\{ \begin{gathered}
  {{\boldsymbol{A}}_k} = \left[ {{{\left( {{\boldsymbol{\hat P}}_{t|t - 1}^{b;{x_2}}} \right)}_k} + {\boldsymbol{F}}_t^{\text{T}}{{\left( {{\boldsymbol{\hat P}}_{t|t - 1}^{b;{x_1}{x_2}}} \right)}_k}} \right], \hfill \\
  {{\boldsymbol{B}}_k} = \left[ {{\boldsymbol{F}}_t^{\text{T}}{{\left( {{\boldsymbol{\hat P}}_{t|t - 1}^{b;{x_1}}} \right)}_k} + {{\left( {{\boldsymbol{\hat P}}_{t|t - 1}^{b;{x_2}{x_1}}} \right)}_k}} \right], \hfill \\
  {\boldsymbol{C}} = {\boldsymbol{I}}, \hfill \\
  {\boldsymbol{D}} = {{\boldsymbol{F}}_t}. \hfill \\ 
\end{gathered}  \right.
\end{align}
${\left[ {{\boldsymbol{W}}_t^{\text{T}}{{\left( {{{{\boldsymbol{\hat \Xi }}}_t}} \right)}_k}{{\boldsymbol{W}}_t}} \right]^{ - 1}}$ can be further written as
\begin{align}\label{nfAKBTjajltb}
{\left[ {{\boldsymbol{W}}_t^{\text{T}}{{\left( {{{{\boldsymbol{\hat \Xi }}}_t}} \right)}_k}{{\boldsymbol{W}}_t}} \right]^{ - 1}} = {\left( {{\boldsymbol{L}}_t^b} \right)_k} = {\boldsymbol{A}}_k^{ - 1} - {\left( {{\boldsymbol{\hat K}}_t^b} \right)_k}{{\boldsymbol{F}}_t}{\boldsymbol{A}}_k^{ - 1}
\end{align}
with
\begin{align}\label{FABJIBAk}
{\left( {{\boldsymbol{\hat K}}_t^b} \right)_k} = {\boldsymbol{A}}_k^{ - 1}{{\boldsymbol{B}}_k}{\left( {{\boldsymbol{I}} + {{\boldsymbol{F}}_t}{\boldsymbol{A}}_k^{ - 1}{{\boldsymbol{B}}_k}} \right)^{ - 1}},
\end{align}
where ${\left( {{\boldsymbol{\hat K}}_t^b} \right)_k}$ is defined as the smoothing gain.

From \eqref{FABJIBAk}, one can obtain ${\left( {{\boldsymbol{\hat K}}_t^b} \right)_k}\left( {{\boldsymbol{I}} + {{\boldsymbol{F}}_t}{\boldsymbol{A}}_k^{ - 1}{{\boldsymbol{B}}_k}} \right){\text{ = }}{\boldsymbol{A}}_k^{ - 1}{{\boldsymbol{B}}_k}$. Multiply ${{\boldsymbol{B}}_k}$ on both sides of \eqref{nfAKBTjajltb}, and we can obtain
\begin{align}\label{31KbtABFK}
\begin{gathered}
  {\left( {{\boldsymbol{L}}_t^b} \right)_k}{{\boldsymbol{B}}_k} = {\boldsymbol{A}}_k^{ - 1}{{\boldsymbol{B}}_k} - {\left( {{\boldsymbol{\hat K}}_t^b} \right)_k}{{\boldsymbol{F}}_t}{\boldsymbol{A}}_k^{ - 1}{{\boldsymbol{B}}_k} \hfill \\
   = {\left( {{\boldsymbol{\hat K}}_t^b} \right)_k} + {\left( {{\boldsymbol{\hat K}}_t^b} \right)_k}{{\boldsymbol{F}}_t}{\boldsymbol{A}}_k^{ - 1}{{\boldsymbol{B}}_k} - {\left( {{\boldsymbol{\hat K}}_t^b} \right)_k}{{\boldsymbol{F}}_t}{\boldsymbol{A}}_k^{ - 1}{{\boldsymbol{B}}_k} \hfill \\
   = {\left( {{\boldsymbol{\hat K}}_t^b} \right)_k}. \hfill \\ 
\end{gathered} 
\end{align}
Combining \eqref{nfAKBTjajltb}, \eqref{DbtStTWbt}, and \eqref{31KbtABFK} leads to the updated strategy of ${{{\boldsymbol{\tilde x}}}_t}$
\begin{align}\label{38xtTxmao}
\begin{gathered}
  {{\boldsymbol{x}}_{t|T}} = {\left[ {{\boldsymbol{W}}_t^{\text{T}}{{\left( {{{{\boldsymbol{\hat \Xi }}}_t}} \right)}_k}{{\boldsymbol{W}}_t}} \right]^{ - 1}}{\boldsymbol{W}}_t^{\text{T}}{\left( {{{{\boldsymbol{\hat \Xi }}}_t}} \right)_k}{{\boldsymbol{\Theta }}_t} \hfill \\
   = {\left( {{\boldsymbol{L}}_t^b} \right)_k}{{\boldsymbol{B}}_k}{{{\boldsymbol{\tilde x}}}_{t + 1|T}} + {\left( {{\boldsymbol{L}}_t^b} \right)_k}{{\boldsymbol{A}}_k}{{{\boldsymbol{\tilde x}}}_{t|t}} \hfill \\
   = {\left( {{\boldsymbol{\hat K}}_t^b} \right)_k}{{{\boldsymbol{\tilde x}}}_{t + 1|T}} + \left[ {{\boldsymbol{A}}_k^{ - 1} - {{\left( {{\boldsymbol{\hat K}}_t^b} \right)}_k}{{\boldsymbol{F}}_t}{\boldsymbol{A}}_k^{ - 1}} \right]{{\boldsymbol{A}}_k}{{{\boldsymbol{\tilde x}}}_{t|t}} \hfill \\
   = {{{\boldsymbol{\tilde x}}}_{t|t}} + {\left( {{\boldsymbol{\hat K}}_t^b} \right)_k}\left( {{{{\boldsymbol{\tilde x}}}_{t + 1|T}} - {{{\boldsymbol{\tilde x}}}_{t + 1|t}}} \right). \hfill \\ 
\end{gathered} 
\end{align}
where ${\left( {{\boldsymbol{\hat K}}_t^b} \right)_k}$ is
\begin{align}\label{gainmeesmoother}
\begin{gathered}
  {\left( {{\boldsymbol{\hat K}}_t^b} \right)_k} = {\left[ {{\boldsymbol{W}}_t^{\text{T}}{{\left( {{{{\boldsymbol{\hat \Xi }}}_t}} \right)}_k}{{\boldsymbol{W}}_t}} \right]^{ - 1}} \hfill \\
   \times \left[ {{\boldsymbol{F}}_t^{\text{T}}{{\left( {{\boldsymbol{\hat P}}_{t|t - 1}^{b;{x_1}}} \right)}_k} + {{\left( {{\boldsymbol{\hat P}}_{t|t - 1}^{b;{x_2}{x_1}}} \right)}_k}} \right]. \hfill \\ 
\end{gathered} 
\end{align}
When condition ${{\left\| {{{\left( {{{{\boldsymbol{\tilde x}}}_{t|T}}} \right)}_{k + 1}} - {{\left( {{{{\boldsymbol{\tilde x}}}_{t|T}}} \right)}_k}} \right\|} \mathord{\left/
 {\vphantom {{\left\| {{{\left( {{{{\boldsymbol{\tilde x}}}_{t|T}}} \right)}_{k + 1}} - {{\left( {{{{\boldsymbol{\tilde x}}}_{t|T}}} \right)}_k}} \right\|} {\left\| {{{\left( {{{{\boldsymbol{\tilde x}}}_{t|T}}} \right)}_k}} \right\|}}} \right.
 \kern-\nulldelimiterspace} {\left\| {{{\left( {{{{\boldsymbol{\tilde x}}}_{t|T}}} \right)}_k}} \right\|}} \leqslant \tau $ is satisfied, the the FPI algorithm stops, and ${\boldsymbol{\hat K}}_t^b = {\left( {{\boldsymbol{\hat K}}_t^b} \right)_k}$.
 
The error covariance of smoothed estimate ${{{\boldsymbol{\tilde x}}}_{t|T}}$ can be obtained by
\begin{align}\label{tKTtPtJ1IjN}
{{\boldsymbol{P}}_{t|T}} = {{\boldsymbol{P}}_{t|t}} + {\boldsymbol{K}}_t^b\left( {{{\boldsymbol{P}}_{t + 1|N}} - {{\boldsymbol{P}}_{t + 1|t}}} \right){\left( {{\boldsymbol{K}}_t^b} \right)^{\text{T}}}.
\end{align}
The detailed derivation of \eqref{tKTtPtJ1IjN} is shown in \ref{deadapin}. 

According to the above derivation, the pseudo-code of the MEE-RTS smoother is summarized in Algorithm \ref{MEERTS-Smoother}.

\begin{algorithm}[t]
\caption{RTS smoother based on the MEE} \label{MEERTS-Smoother}
\LinesNumbered 
\KwIn{state transition matrix ${\boldsymbol{F}_t}$, measurement matrix ${\boldsymbol{H}_t}$, covariance matrices ${\boldsymbol{Q}_t}$, ${\boldsymbol{R}_t}$, and kernel bandwidth $\sigma$}
\KwOut{${{\boldsymbol{\tilde x}}_{t|T}}$}
\textbf{Initialization}: ${{\boldsymbol{\tilde x}}_{0|0}} = {{\boldsymbol{x}}_0}$, ${{\boldsymbol{P}}_{0|0}} = {{\boldsymbol{P}}_0}$\; 
\textbf{Forward Pass:}\
\For{$t\leftarrow $1$ $ \KwTo $T$}{
Calculate ${{\boldsymbol{\tilde x}}_{t|t - 1}}$ and ${{\boldsymbol{P}}_{t|t - 1}}$ using 
\begin{subequations}\label{prespttj1dqt}
\begin{numcases}{}
{{{\boldsymbol{\tilde x}}}_{t|t - 1}} = {{\boldsymbol{F}}_t}{{{\boldsymbol{\tilde x}}}_{t - 1|t - 1}},\\\label{tr71qkj1axkjej} 
{{\boldsymbol{P}}_{t|t - 1}} = {{\boldsymbol{F}}_t}{{\boldsymbol{P}}_{t - 1|t - 1}}{\boldsymbol{F}}_t^{\text{T}} + {{\boldsymbol{Q}}_t};\label{vkjxkCdYk}
\end{numcases}
\end{subequations}

\While{${{\left\| {{{\left( {{{{\boldsymbol{\tilde x}}}_{t|t}}} \right)}_{k + 1}} - {{\left( {{{{\boldsymbol{\tilde x}}}_{t|t}}} \right)}_k}} \right\|} \mathord{\left/
 {\vphantom {{\left\| {{{\left( {{{{\boldsymbol{\tilde x}}}_{t|t}}} \right)}_{k + 1}} - {{\left( {{{{\boldsymbol{\tilde x}}}_{t|t}}} \right)}_k}} \right\|} {\left\| {{{\left( {{{{\boldsymbol{\tilde x}}}_{t|t}}} \right)}_k}} \right\|}}} \right.
 \kern-\nulldelimiterspace} {\left\| {{{\left( {{{{\boldsymbol{\tilde x}}}_{t|t}}} \right)}_k}} \right\|}} \geqslant \tau $}{Update ${{{\left( {{{{\boldsymbol{\tilde x}}}_{t|t}}} \right)}_{k + 1}}}$ using \eqref{xtzTjTddeyTe}, \eqref{outfaizhzTjtfaizh}, \eqref{PPPPxy}, \eqref{XtKtYtjHx}, and \eqref{KgainKj1yjZoTzK};}
Update ${{\boldsymbol{P}}_{t|t}}$ using \eqref{pTtIrKtTjTjIdanwei}\;
}
\textbf{Backward Pass:}\
\For{$t\leftarrow $T - 1$ $ \KwTo $1$}{
Calculate ${{{{\boldsymbol{\tilde x}}}_{t + 1|t}}}$ and ${{{{\boldsymbol{P}}_{t + 1|t}}}}$ utilizing \eqref{prespttj1dqt}\;
Let $k=1$ and  ${\left( {{{{\boldsymbol{\tilde x}}}_{t|T}}} \right)_{k = 0}} = {{{\boldsymbol{\tilde x}}}_{t|t}}$\;
\While{${{\left\| {{{\left( {{{{\boldsymbol{\tilde x}}}_{t|T}}} \right)}_{k + 1}} - {{\left( {{{{\boldsymbol{\tilde x}}}_{t|T}}} \right)}_k}} \right\|} \mathord{\left/
 {\vphantom {{\left\| {{{\left( {{{{\boldsymbol{\tilde x}}}_{t|T}}} \right)}_{k + 1}} - {{\left( {{{{\boldsymbol{\tilde x}}}_{t|T}}} \right)}_k}} \right\|} {\left\| {{{\left( {{{{\boldsymbol{\tilde x}}}_{t|T}}} \right)}_k}} \right\|}}} \right.
 \kern-\nulldelimiterspace} {\left\| {{{\left( {{{{\boldsymbol{\tilde x}}}_{t|T}}} \right)}_k}} \right\|}} \geqslant \tau $}{
Update ${\left( {{{{\boldsymbol{\tilde x}}}_{t|T}}} \right)_{k + 1}}$ using \eqref{38xtTxmao}
with \eqref{gainmeesmoother}, \eqref{ssssttt11122122}, \eqref{27PPPPx12}, and \eqref{thetajwtxt};
}
Calculate ${{\boldsymbol{P}}_{t|T}}$ utilizing \eqref{tKTtPtJ1IjN}\;
}

\end{algorithm}

\begin{algorithm}[t]
\caption{Extended RTS smoother based on the MEE} \label{MEERTS-ExtendedSmoother}
\LinesNumbered 
\KwIn{${\boldsymbol{f}}\left(  \cdot  \right)$, ${\boldsymbol{h}}\left(  \cdot  \right)$, ${\boldsymbol{Q}_t}$, ${\boldsymbol{R}_t}$, and $\sigma$}
\KwOut{${{\boldsymbol{\tilde x}}_{t|T}}$}
\textbf{Initialization}: ${{\boldsymbol{\tilde x}}_{0|0}} = {{\boldsymbol{x}}_0}$, ${{\boldsymbol{P}}_{0|0}} = {{\boldsymbol{P}}_0}$\; 
\textbf{Forward Pass:}\
\For{$t\leftarrow $1$ $ \KwTo $T$}{
These steps of the MEE-EKF algorithm are presented in \cite{chen2019minimum}. 
}
\textbf{Backward Pass:}\
\For{$t\leftarrow $T - 1$ $ \KwTo $1$}{
Predicting the state of the system using ${{\boldsymbol{\tilde x}}_{t + 1|t}} = {\boldsymbol{f}}\left( {{{{\boldsymbol{\tilde x}}}_{t|t}}} \right)$\;
Calculate the Jacobian matrix and covariance matrix using \eqref{yjkbjuzhenF} and ${{\boldsymbol{P}}_{t|t - 1}} = {{\boldsymbol{F}}_{t - 1}}{{\boldsymbol{P}}_{t - 1|t - 1}}{\boldsymbol{F}}_{t - 1}^{\text{T}} + {\boldsymbol{Q}}$ \;
Let $k=1$ and  ${\left( {{{{\boldsymbol{\tilde x}}}_{t|T}}} \right)_{k = 0}} = {{{\boldsymbol{\tilde x}}}_{t|t}}$\;
\While{${{\left\| {{{\left( {{{{\boldsymbol{\tilde x}}}_{t|T}}} \right)}_{k + 1}} - {{\left( {{{{\boldsymbol{\tilde x}}}_{t|T}}} \right)}_k}} \right\|} \mathord{\left/
 {\vphantom {{\left\| {{{\left( {{{{\boldsymbol{\tilde x}}}_{t|T}}} \right)}_{k + 1}} - {{\left( {{{{\boldsymbol{\tilde x}}}_{t|T}}} \right)}_k}} \right\|} {\left\| {{{\left( {{{{\boldsymbol{\tilde x}}}_{t|T}}} \right)}_k}} \right\|}}} \right.
 \kern-\nulldelimiterspace} {\left\| {{{\left( {{{{\boldsymbol{\tilde x}}}_{t|T}}} \right)}_k}} \right\|}} \geqslant \tau $}{
Update ${\left( {{{{\boldsymbol{\tilde x}}}_{t|T}}} \right)_{k + 1}}$ using ${\left( {{{{\boldsymbol{\tilde x}}}_{t|T}}} \right)_{k + 1}} = {{\boldsymbol{\tilde x}}_{t|t}} + {\boldsymbol{\hat K}}_t^b\left( {{{{\boldsymbol{\tilde x}}}_{t + 1|T}} - {{{\boldsymbol{\tilde x}}}_{t + 1|t}}} \right)$
with $\left\{ \begin{gathered}
  {\boldsymbol{\hat K}}_t^b = {\left( {{\boldsymbol{W}}_t^{\text{T}}{{{\boldsymbol{\hat \Xi }}}_t}{{\boldsymbol{W}}_t}} \right)^{ - 1}}\left( {{\boldsymbol{F}}_t^{\text{T}}{\boldsymbol{\hat P}}_{t|t - 1}^{b;{x_1}} + {\boldsymbol{\hat P}}_{t|t - 1}^{b;{x_2}{x_1}}} \right), \hfill \\
  {{{\boldsymbol{\hat \Xi }}}_t} = {\boldsymbol{\hat \Gamma }}_t^{\text{T}}{{{\boldsymbol{\hat \Gamma }}}_t} + {\boldsymbol{\hat \Lambda }}_t^{\text{T}}{{{\boldsymbol{\hat \Lambda }}}_t},{\left[ {{{{\boldsymbol{\hat \Lambda }}}_t}} \right]_{ij}} = {\text{G}}\left[ {\hat \varepsilon _{t;i}^l - \hat \varepsilon _{t;j}^l} \right], \hfill \\
  {\left[ {{{{\boldsymbol{\hat \Gamma }}}_t}} \right]_{ij}} = \left\{ {\begin{array}{*{20}{l}}
  {\sum\limits_{j = 1}^L {{{\text{G}}_\sigma }\left[ {\hat \varepsilon _{t;i}^l - \hat \varepsilon _{t;j}^l} \right],i = j,} } \\ 
  {0,i \ne j,} 
\end{array}} \right. \hfill \\
  {\boldsymbol{\hat P}}_{t|t - 1}^{b;{x_1}} = {\left( {{\boldsymbol{Q}}_t^{ - \tfrac{1}{2}}} \right)^{\text{T}}}{{{\boldsymbol{\hat \Xi }}}_{t;{x_1}}}{\boldsymbol{Q}}_t^{ - \tfrac{1}{2}}, \hfill \\
  {{\hat \varepsilon }_{t;i}} = {\theta _{t;i}} - {{\boldsymbol{w}}_{t;i}}{\left( {{{{\boldsymbol{\tilde x}}}_{t|T}}} \right)_k}, \hfill \\
  {\boldsymbol{\hat P}}_{t|t - 1}^{b;{x_2}{x_1}} = {\left( {{\boldsymbol{P}}_{t|t}^{ - \tfrac{1}{2}}} \right)^{\text{T}}}{{{\boldsymbol{\hat \Xi }}}_{t;{x_2}{x_1}}}{\boldsymbol{Q}}_t^{ - \tfrac{1}{2}}, \hfill \\ 
\end{gathered}  \right.$
}
Update covariance matrix using 
\begin{align}
{{\boldsymbol{P}}_{t|T}} \approx {{\boldsymbol{P}}_{t|t}} + {\boldsymbol{\hat K}}_t^b\left( {{{\boldsymbol{P}}_{t + 1|N}} - {{\boldsymbol{P}}_{t + 1|t}}} \right){\left( {{\boldsymbol{\hat K}}_t^b} \right)^{\text{T}}}
\end{align}
}
\end{algorithm}

\subsection{The MEE based extended RTS smoother}
For state estimation of the nonlinear systems, the process and measurement equations can be described as
\begin{align}\label{nonlinear}
\left\{ \begin{gathered}
  {{\boldsymbol{x}}_t} = {\boldsymbol{f}}\left( {{{\boldsymbol{x}}_{t - 1}}} \right) + {{\boldsymbol{q}}_{t - 1}}, \hfill \\
  {{\boldsymbol{y}}_t} = {\boldsymbol{h}}\left( {{{\boldsymbol{x}}_t}} \right) + {{\boldsymbol{r}}_t}. \hfill \\ 
\end{gathered}  \right.
\end{align}
where mapping ${\boldsymbol{f}}\left(  \cdot  \right)$ denotes a known nonlinear state transition function, mapping ${\boldsymbol{h}}\left(  \cdot  \right)$ is a known nonlinear measurement function. The ERTS is a frequently utilized tool for the state smoothing problems for the nonlinear systems. In the ERTS, the nonlinear functions ${\boldsymbol{f}}\left(  \cdot  \right)$ is linearized by the first-order Taylor series expansion at ${{{{\boldsymbol{\tilde x}}}_{t - 1|t - 1}}}$, i.e.,
\begin{align}\label{34Htxjxmtstj1}
{\boldsymbol{f}}\left( {{{\boldsymbol{x}}_{t - 1}}} \right) \approx {\boldsymbol{f}}\left( {{{{\boldsymbol{\tilde x}}}_{t - 1|t - 1}}} \right) + {{\boldsymbol{F}}_{t - 1}}\left( {{{\boldsymbol{x}}_{t - 1}} - {{{\boldsymbol{\tilde x}}}_{t - 1|t - 1}}} \right),
\end{align}
where the Jacobian matrix of the nonlinear function ${\boldsymbol{f}}\left(  \cdot  \right)$ is obtained via
\begin{align}\label{yjkbjuzhenF}
{{\boldsymbol{F}}_{t - 1}} = {\left. {\frac{{\partial {\boldsymbol{f}}\left( {{{\boldsymbol{x}}_{t - 1}}} \right)}}{{\partial {\boldsymbol{x}}}}} \right|_{{\boldsymbol{x}} = {{{\boldsymbol{\tilde x}}}_{t - 1|t - 1}}}}.
\end{align}

Combining \eqref{nonlinear} and \eqref{34Htxjxmtstj1} yields:
\begin{align}\label{xqxhhy}
{{\boldsymbol{x}}_t} \approx {\boldsymbol{f}}\left( {{{{\boldsymbol{\tilde x}}}_{t - 1|t - 1}}} \right) + {{\boldsymbol{F}}_{t - 1}}\left( {{{\boldsymbol{x}}_{t - 1}} - {{{\boldsymbol{\tilde x}}}_{t - 1|t - 1}}} \right) + {{\boldsymbol{q}}_{t - 1}}.
\end{align}

Substitute \eqref{xqxhhy} into \eqref{nonlinear} and the augmented model for the MEE-ERTS can be written as
\begin{align}
\begin{gathered}
  {\boldsymbol{\varepsilon }}_t^l = \left[ {\begin{array}{*{20}{c}}
  {{\boldsymbol{Q}}_t^{ - \tfrac{1}{2}}}&{{{\boldsymbol{0}}_n}} \\ 
  {{{\boldsymbol{0}}_n}}&{{\boldsymbol{P}}_{t|t}^{ - \tfrac{1}{2}}} 
\end{array}} \right]\left[ {\begin{array}{*{20}{c}}
  {{{{\boldsymbol{\tilde x}}}_{t + 1|T}} - {\boldsymbol{f}}\left( {{{\boldsymbol{x}}_t}} \right)} \\ 
  {{{{\boldsymbol{\tilde x}}}_{t|t}} - {{\boldsymbol{x}}_t}} 
\end{array}} \right] \hfill \\
   = \left[ {\begin{array}{*{20}{c}}
  {{\boldsymbol{Q}}_t^{ - \tfrac{1}{2}}}&{{{\boldsymbol{0}}_n}} \\ 
  {{{\boldsymbol{0}}_n}}&{{\boldsymbol{P}}_{t|t}^{ - \tfrac{1}{2}}} 
\end{array}} \right]\left[ {\begin{array}{*{20}{c}}
  \begin{gathered}
  {{{\boldsymbol{\tilde x}}}_{t + 1|T}} - {\boldsymbol{f}}\left( {{{{\boldsymbol{\tilde x}}}_{t|t}}} \right) -  \hfill \\
  {{\boldsymbol{F}}_t}\left( {{{\boldsymbol{x}}_t} - {{{\boldsymbol{\tilde x}}}_{t|t}}} \right) \hfill \\ 
\end{gathered}  \\ 
  {{{{\boldsymbol{\tilde x}}}_{t|t}} - {{\boldsymbol{x}}_t}} 
\end{array}} \right]. \hfill \\ 
\end{gathered} 
\end{align}
and we also obtain
\begin{align}\label{DbEtWbExtjebt}
{\boldsymbol{\hat \varepsilon }}_t^l = {\boldsymbol{\Theta }}_t^l - {\boldsymbol{W}}_t^l{\left( {{{{\boldsymbol{\tilde x}}}_{t|t}}} \right)_k}
\end{align}
with
\begin{align}\label{47xmaottgPje21}
{\boldsymbol{\Theta }}_t^l = \left[ {\begin{array}{*{20}{c}}
  {{\boldsymbol{Q}}_t^{ - \tfrac{1}{2}}\left[ {{{{\boldsymbol{\tilde x}}}_{t + 1|T}} - {\boldsymbol{f}}\left( {{{{\boldsymbol{\tilde x}}}_{t|t}}} \right) + {{\boldsymbol{F}}_t}{{{\boldsymbol{\tilde x}}}_{t|t}}} \right]} \\ 
  {{\boldsymbol{P}}_{t|t}^{ - \tfrac{1}{2}}{{{\boldsymbol{\tilde x}}}_{t|t}}} 
\end{array}} \right]
\end{align}
and
\begin{align}\label{4884pttgfue12jP}
{\boldsymbol{W}}_t^l = \left[ {\begin{array}{*{20}{c}}
  {{\boldsymbol{Q}}_t^{ - \tfrac{1}{2}}{{\boldsymbol{F}}_t}} \\ 
  {{\boldsymbol{P}}_{t|t}^{ - \tfrac{1}{2}}} 
\end{array}} \right].
\end{align}
It is apparent that \eqref{DbEtWbExtjebt} have a similar form as the one in \eqref{thetajwtxt}. Hence, the derivation process of the backward pass of the MEE-ERTS smoother is similar to that of the backward pass of the MEE-RTS smoother. In order to avoid the repetition, the derivation process of the MEE-ERTS smoother is omitted and its pseudo-code is presented in Algorithm \ref{MEERTS-ExtendedSmoother}.

\begin{remark}
It is notable that the threshold $\tau$, in both Algorithm \ref{MEERTS-Smoother} and Algorithm \ref{MEERTS-ExtendedSmoother}, is a vital parameter to control the number of the FPI method. The larger the threshold is, the fewer the number of FPI are required, and therefore the less computational burden of the algorithms, which in turn reduces the performance of the proposed smoothers in steady-state errors to some extent, whereas a smaller threshold means the opposite. 
\end{remark}

\section{Theoretical analysis}\label{tanluysis}
\subsection{Mean error behavior}
In this part, ${{\boldsymbol{\xi }}_{s;t}} = {{\boldsymbol{x}}_t} - {{\boldsymbol{\tilde x}}_{t|T}}$ is employed to present the estimation error of smoother, and ${{\boldsymbol{\xi }}_{f;t}} = {{\boldsymbol{x}}_t} - {{\boldsymbol{\tilde x}}_{t|t}}$ denotes the estimation error of filter. According to \eqref{31KbtABFK}, ${{\boldsymbol{\xi }}_{s;t}}$ can be rewritten as:
\begin{align}\label{kyplslongktb}
{{\boldsymbol{\xi }}_{s;t}} &= {{\boldsymbol{x}}_t} - {{{\boldsymbol{\tilde x}}}_{t|T}}\notag\\
 &= {{\boldsymbol{\xi }}_{f;t}} - {\boldsymbol{\hat K}}_t^b\left( {{{{\boldsymbol{\tilde x}}}_{t + 1|T}} - {{{\boldsymbol{\tilde x}}}_{t + 1|t}}} \right).
\end{align}
According to ${{{\boldsymbol{\tilde x}}}_{t|t - 1}} = {{\boldsymbol{F}}_t}{{{\boldsymbol{\tilde x}}}_{t - 1|t - 1}}$, ${{\boldsymbol{\xi }}_{f;t - 1}} = {{\boldsymbol{x}}_{t - 1}} - {{{\boldsymbol{\tilde x}}}_{t - 1|t - 1}}$, and ${{\boldsymbol{x}}_t} = {{\boldsymbol{F}}_t}{{\boldsymbol{x}}_{t - 1}} + {{\boldsymbol{q}}_t}$, one can obtain
\begin{align}\label{51sftjj1jianQ2}
{{{\boldsymbol{\tilde x}}}_{t + 1|t}} = {{\boldsymbol{x}}_{t + 1}} - {{\boldsymbol{q}}_{t + 1}} - {{\boldsymbol{F}}_t}{{\boldsymbol{\xi }}_{f;t}}.
\end{align}
Substituting \eqref{51sftjj1jianQ2} into \eqref{kyplslongktb} yields 
\begin{align}\label{50qtjekbtssa}
{{\boldsymbol{\xi }}_{s;t}} = \left( {{\boldsymbol{I}} - {\boldsymbol{\hat K}}_t^b{{\boldsymbol{F}}_t}} \right){{\boldsymbol{\xi }}_{f;t}} + {\boldsymbol{\hat K}}_t^b{{\boldsymbol{\xi }}_{s;t + 1}} - {\boldsymbol{\hat K}}_t^b{{\boldsymbol{q}}_{t + 1}}.
\end{align}
Since the process and measurement noises are zero mean, the expectation of estimation error ${{\boldsymbol{\xi }}_{s;t}}$ can be expressed as
\begin{align}\label{51sanstj1fenha}
{\text{E}}\left[ {{{\boldsymbol{\xi }}_{s;t}}} \right] = \left( {{\boldsymbol{I}} - {\boldsymbol{\hat K}}_t^b{{\boldsymbol{F}}_t}} \right){\text{E}}\left[ {{{\boldsymbol{\xi }}_{f;t}}} \right]{\text{ + }}{\boldsymbol{\hat K}}_t^b{\text{E}}\left[ {{{\boldsymbol{\xi }}_{s;t + 1}}} \right].
\end{align}
From \eqref{27PPPPx12}, it is obvious that ${{\boldsymbol{\hat P}}_{t|t - 1}^{b;{x_1}}}$ and ${{\boldsymbol{\hat P}}_{t|t - 1}^{b;{x_2}}}$ are both strictly diagonally dominant. One can obtain
\begin{align}\label{52KbtFyudengyu}
{\boldsymbol{\hat K}}_t^b{{\boldsymbol{F}}_t} \approx {\left( {{\boldsymbol{F}}_t^{\text{T}}{\boldsymbol{\hat P}}_{t|t - 1}^{b;{x_1}}{{\boldsymbol{F}}_t} + {\boldsymbol{\hat P}}_{t|t - 1}^{b;{x_2}}} \right)^{ - 1}}\left( {{\boldsymbol{F}}_t^{\text{T}}{\boldsymbol{\hat P}}_{t|t - 1}^{b;{x_1}}{{\boldsymbol{F}}_t}} \right).
\end{align}
Substituting \eqref{52KbtFyudengyu} into \eqref{51sanstj1fenha} yields 
\begin{align}\label{Essfhtyudyfni}
\begin{gathered}
  {\text{E}}\left[ {{{\boldsymbol{\xi }}_{s;t}}} \right] \approx \left[ {{\boldsymbol{I}} - {{\left( {{\boldsymbol{F}}_t^{\text{T}}{\boldsymbol{\hat P}}_{t|t - 1}^{b;{x_1}}{{\boldsymbol{F}}_t} + {\boldsymbol{\hat P}}_{t|t - 1}^{b;{x_2}}} \right)}^{ - 1}}\left( {{\boldsymbol{F}}_t^{\text{T}}{\boldsymbol{\hat P}}_{t|t - 1}^{b;{x_1}}{{\boldsymbol{F}}_t}} \right)} \right] \hfill \\
   \times {\text{E}}\left[ {{{\boldsymbol{\xi }}_{f;t}}} \right] + \left[ {{{\left( {{\boldsymbol{F}}_t^{\text{T}}{\boldsymbol{\hat P}}_{t|t - 1}^{b;{x_1}}{{\boldsymbol{F}}_t} + {\boldsymbol{\hat P}}_{t|t - 1}^{b;{x_2}}} \right)}^{ - 1}}\left( {{\boldsymbol{F}}_t^{\text{T}}{\boldsymbol{\hat P}}_{t|t - 1}^{b;{x_1}}{{\boldsymbol{F}}_t}} \right)} \right] \times  \hfill \\
  {\boldsymbol{F}}_t^{ - 1}{\text{E}}\left[ {{{\boldsymbol{\xi }}_{s;t + 1}}} \right]. \hfill \\ 
\end{gathered} 
\end{align}
From \eqref{Essfhtyudyfni}, one can obtain that ${\text{E}}\left[ {{{\boldsymbol{\xi }}_{s;t}}} \right]$ is the linear combination of ${\text{E}}\left[ {{{\boldsymbol{\xi }}_{f;t}}} \right]$ and ${\text{E}}\left[ {{{\boldsymbol{\xi }}_{s;t + 1}}} \right]$. Assuming that ${{\boldsymbol{F}}_t}$ is stable, one can conclude that both ${\boldsymbol{F}}_t^{\text{T}}{\boldsymbol{\hat P}}_{t|t - 1}^{b;{x_1}}{{\boldsymbol{F}}_t}$ and ${\boldsymbol{\hat P}}_{t|t - 1}^{b;{x_2}}$ are positive semi-definite, which will lead to the fact that \eqref{52KbtFyudengyu} is stable. Further, \eqref{Essfhtyudyfni} is also stable, which means that the proposed MEE-RTS smoother operates stably.

\subsection{Mean square error behavior}
According to \eqref{50qtjekbtssa}, the error covariance matrix of the state can be written as
\begin{align}\label{EsansanQFBT}
\begin{gathered}
  {\text{E}}\left[ {{{\boldsymbol{\xi }}_{s;t}}{\boldsymbol{\xi }}_{s;t}^{\text{T}}} \right] =  \hfill \\
  \left( {{\boldsymbol{I}} - {\text{E}}\left[ {{\boldsymbol{\hat K}}_t^b} \right]{{\boldsymbol{F}}_t}} \right){\text{E}}\left[ {{{\boldsymbol{\xi }}_{f;t}}{\boldsymbol{\xi }}_{f;t}^{\text{T}}} \right]{\left( {{\boldsymbol{I}} - {\text{E}}\left[ {{\boldsymbol{\hat K}}_t^b} \right]{{\boldsymbol{F}}_t}} \right)^{\text{T}}} \hfill \\
   + {\text{E}}\left[ {{\boldsymbol{\hat K}}_t^b} \right]{\text{E}}\left[ {{{\boldsymbol{\xi }}_{s;t + 1}}{\boldsymbol{\xi }}_{s;t + 1}^{\text{T}}} \right]{\text{E}}\left[ {{{\left( {{\boldsymbol{\hat K}}_t^b} \right)}^{\text{T}}}} \right] \hfill \\
  {\text{ + E}}\left[ {{\boldsymbol{\hat K}}_t^b} \right]{{\boldsymbol{Q}}_t}{\text{E}}\left[ {{{\left( {{\boldsymbol{\hat K}}_t^b} \right)}^{\text{T}}}} \right]. \hfill \\ 
\end{gathered} 
\end{align}
A convergent gain of smoother ${\boldsymbol{K}}_t^b$ can be obtained utilizing the fixed-point algorithm, and the expectation of gain ${\text{E}}\left[ {{\boldsymbol{K}}_t^b} \right]$ can be obtained by using the adaptive solution
\begin{align}
{\text{E}}\left[ {{\boldsymbol{K}}_t^b} \right] = \left( {1 - \iota } \right){\text{E}}\left[ {{\boldsymbol{K}}_{t - 1}^b} \right] + \iota {\boldsymbol{K}}_t^b,
\end{align}
where $\iota $ is the forgotten factor. 

Formula \eqref{EsansanQFBT} can be simplified as
\begin{align}\label{56NtOtOTt}
{{\boldsymbol{{\rm N}}}_t}{\text{ = }}{{\boldsymbol{\Omega }}_t}{{\boldsymbol{{\rm N}}}_{t - 1}}{\boldsymbol{\Omega }}_t^{\text{T}}{\text{ + }}{{\boldsymbol{Y}}_t}
\end{align}
with
\begin{align}
\left\{ \begin{gathered}
  {{\boldsymbol{{\rm N}}}_t} = {\text{E}}\left[ {{{\boldsymbol{\xi }}_{s;t}}{\boldsymbol{\xi }}_{s;t}^{\text{T}}} \right], \hfill \\
  {{\boldsymbol{\Omega }}_t} = {\text{E}}\left[ {{\boldsymbol{\hat K}}_t^b} \right], \hfill \\
  {{\boldsymbol{Y}}_t} = \left( {{\boldsymbol{I}} - {\text{E}}\left[ {{\boldsymbol{\hat K}}_t^b} \right]{{\boldsymbol{F}}_t}} \right){\text{E}}\left[ {{{\boldsymbol{\xi }}_{f;t}}{\boldsymbol{\xi }}_{f;t}^{\text{T}}} \right]{\left( {{\boldsymbol{I}} - {\text{E}}\left[ {{\boldsymbol{\hat K}}_t^b} \right]{{\boldsymbol{F}}_t}} \right)^{\text{T}}} \hfill \\
  {\text{ + E}}\left[ {{\boldsymbol{\hat K}}_t^b} \right]{{\boldsymbol{Q}}_t}{\text{E}}\left[ {{{\left( {{\boldsymbol{\hat K}}_t^b} \right)}^{\text{T}}}} \right]. \hfill \\ 
\end{gathered}  \right.
\end{align}\
From the matrix vector operator:
\begin{align}\label{vectorpoerator}
\left\{ \begin{gathered}
  {\text{vec}}\left( {{\boldsymbol{SUV}}} \right) = \left( {{{\boldsymbol{V}}^{\text{T}}} \otimes {\boldsymbol{S}}} \right){\text{vec}}\left( {\boldsymbol{U}} \right), \hfill \\
  {\text{vec}}\left( {{\boldsymbol{S}} + {\boldsymbol{V}}} \right) = {\text{vec}}\left( {\boldsymbol{S}} \right) + {\text{vec}}\left( {\boldsymbol{V}} \right), \hfill \\ 
\end{gathered}  \right.
\end{align}
where $ \otimes $ represents the Kronecker product and ${\text{vec}}\left(  \cdot  \right)$ denotes vectorization operation.

The closed-form solution of \eqref{56NtOtOTt} can be written as
\begin{align}
\mathop {\lim }\limits_{t \to \infty } {\text{vec}}\left( {{{\boldsymbol{{\rm N}}}_t}} \right) = \mathop {\lim }\limits_{t \to \infty } {\left( {{\boldsymbol{I}} - {{\boldsymbol{\Omega }}_t} \otimes {{\boldsymbol{\Omega }}_t}} \right)^{ - 1}}{\text{vec}}\left( {{{\boldsymbol{Y}}_t}} \right),
\end{align}
where ${\text{vec}}\left( {{{\boldsymbol{Y}}_t}} \right)$ can be calculated by using \eqref{vectorpoerator}, and one can be obtain
\begin{align}\label{QVECtSEgZZTFtS}
{\text{vec}}\left( {{{\boldsymbol{Y}}_t}} \right) = {{\boldsymbol{\xi }}_t}{\text{vec}}\left[ {{\text{E}}\left[ {{{\boldsymbol{\xi }}_{f;t}}{\boldsymbol{\xi }}_{f;t}^{\text{T}}} \right]} \right] + {{\boldsymbol{\varsigma }}_t}{\text{vec}}\left( {{{\boldsymbol{Q}}_t}} \right)
\end{align}
with
\begin{align}
\left\{ \begin{gathered}
  {{\boldsymbol{\xi }}_t}{\text{ = }}\left( {{\boldsymbol{I}} - {\text{E}}\left[ {{\boldsymbol{\hat K}}_t^b} \right]{{\boldsymbol{F}}_t}} \right) \otimes \left( {{\boldsymbol{I}} - {\text{E}}\left[ {{\boldsymbol{\hat K}}_t^b} \right]{{\boldsymbol{F}}_t}} \right), \hfill \\
  {{\boldsymbol{\varsigma }}_t} = {\text{E}}\left[ {{\boldsymbol{\hat K}}_t^b} \right] \otimes {\text{E}}\left[ {{\boldsymbol{\hat K}}_t^b} \right]. \hfill \\ 
\end{gathered}  \right.
\end{align}
The error covariance matrix ${{\text{E}}\left[ {{{\boldsymbol{\xi }}_{f;t}}{\boldsymbol{\xi }}_{f;t}^{\text{T}}} \right]}$, in \eqref{QVECtSEgZZTFtS}, has been calculated in \cite{wang2021numerically}.
\subsection{Computational complexity}
In this section, the computational burden of the proposed MEE-RTS smoother is compared with that of the RTS smoother and MC-RTS smoother for the linear systems. It is noted that the MC-RTS smoother in \cite{wang2020maximum} is developed to estimate the state of the nonlinear systems, and the scheme for using MC-RTS smoother to handle state estimation of linear systems is not given. The pseudo-code for the MC-RTS smoother used to estimate the state of the linear systems (MC-RTSL) is summarised in Algorithm \ref{MCCRTS-linear}, which is used as the baseline in terms of computational burden and steady-state error. In Algorithm \ref{MCCRTS-linear}, ${{{\boldsymbol{K}}_{t;MC}}}$ and ${\boldsymbol{K}}_{t;MC}^b$ are gains of forward and backward pass. 

From Algorithm \ref{MCCRTS-linear}, the computational burden of the forward pass and backward pass of the MC-RTSL smoother are 
\begin{align}
\left\{ \begin{gathered}
  {S_{MC\_F}} = 8{n^3} + 10{n^2}m + 6n{m^2} + 2{m^2} - {n^2} \hfill \\
   + 4mn - n + m + 10 + 3O\left( {{m^3}} \right) + O\left( {{n^3}} \right), \hfill \\
  {S_{MC\_B}} = 12{n^3} + 7{n^2} + n + 10 + 4O\left( {{n^3}} \right), \hfill \\ 
\end{gathered}  \right.
\end{align}
where ${S_{MC\_F}}$ and ${S_{MC\_B}}$ represent the computational complexity of the forward pass and backward pass. Hence, the computational complexity of MC-RTSL smoother is
\begin{align}\label{ccmc-rtslbn3}
\begin{gathered}
  {S_{MC - RTSL}} = 20{n^3} + 10{n^2}m + 6n{m^2} + 2{m^2} +  \hfill \\
  6{n^2} + 4mn + m + 20 + 5O\left( {{n^3}} \right) + 3O\left( {{m^3}} \right). \hfill \\ 
\end{gathered} 
\end{align}

The computational complexity of the forward pass of the MEE-RTS smoother is
\begin{align}\label{meeforwardcb}
\begin{gathered}
  {S_{R - MEEKF}} = \left( {7{M_f} + 8} \right){n^3} + 7{M_f}{m^3} - {n^2} \hfill \\
   + \left( {19{M_f} + 6} \right){n^2}m + \left( {15{M_f} + 2} \right)n{m^2} \hfill \\
   + {M_f}O\left( {{n^3}} \right) + {M_f}m + \left( {5{M_f} - 1} \right)n \hfill \\
   + \left( {7{M_f} - 1} \right)nm + 3{M_f}O\left( {{m^3}} \right). \hfill \\ 
\end{gathered}   
\end{align}
where ${{M_f}}$ is the number of the FPI method of the forward pass. The main formulas for the backward pass and their computational complexity are shown in Table \ref{ccotmeerts}, and the computational burden of the backward pass of the MEE-RTS smoother is
\begin{align}\label{meeforwardbp}
\begin{gathered}
  {S_{MEE\_B}} = \left( {48{M_b} + 8} \right){n^3} + \left( {1 - 6{M_b}} \right){n^2} \hfill \\
   + \left( {1 - 9{M_b}} \right)n + 3{M_b}O({n^3}) + 24{M_b}, \hfill \\ 
\end{gathered} 
\end{align}
where ${M_b}$ denotes the number of the FPI method of the backward pass. Combining \eqref{meeforwardcb} and \eqref{meeforwardbp}, the computational complexity of the MEE-RTS smoother is
\begin{align}\label{mee-rtsccb24m3}
\begin{gathered}
  {S_{MEE - RTS}} = \left( {7{M_f} + 48{M_b} + 16} \right){n^3} + 7{M_f}{m^3} \hfill \\
   + \left( {19{M_f} + 6} \right){n^2}m - 6{M_b}{n^2} + \left( {15{M_f} + 2} \right)n{m^2} \hfill \\
   + \left( {{M_f} + 3{M_b}} \right)O\left( {{n^3}} \right) + {M_f}m + \left( {5{M_f} - 9{M_b}} \right)n \hfill \\
   + \left( {7{M_f} - 1} \right)nm + 3{M_f}O\left( {{m^3}} \right) + 24{M_b}. \hfill \\ 
\end{gathered} 
\end{align}

From \eqref{ccmc-rtslbn3} and \eqref{mee-rtsccb24m3}, we can conclude that the computational complexity of the MEE-RTS smoother is slightly higher than that of MC-RTSL smoother. The main reasons for this phenomenon are the adoption of the MEE criterion and the FPI method of the R-MEEKF algorithm. In the MEE-RTS smoother, the MEE criterion is introduced into the forward and backward pass, and the MEE criterion with double summation is more computationally expensive than the MCC criterion with a single summation. The increase in computational burden due to the introduction of the MEE is inevitable. In addition, the computational burden of the FPI method is higher than or equal to the computational complexity of ARM. 

From the above discussion, one can be deduced that reducing the number of the FPI method (the number can be controlled by the parameter $\tau$ in \cite{chen2019minimum}) or using ARM can significantly lower the computational complexity of the MEE-RTS algorithm. However, these two solutions also decrease the performance of the MEE-RTS smoother in steady-state error. Hence, the choice of parameter $\tau$ should therefore trade-off the steady-state error and the computational complexity of the MEE-RTS smoother, the detailed selection scheme of the parameter $\tau$ is present in Section \ref{para_perfro_mee}.

\begin{algorithm}[t]
\caption{MC-RTS smoother for linear systems} \label{MCCRTS-linear}
\LinesNumbered 
\KwIn{state transition matrix ${\boldsymbol{F}}$, measurement matrix ${\boldsymbol{H}}$, covariance matrices ${\boldsymbol{Q}}$, ${\boldsymbol{R}}$, and kernel bandwidth $\sigma$}
\KwOut{${{\boldsymbol{\tilde x}}_{t|T}}$}
\textbf{Initialization}: ${{\boldsymbol{\tilde x}}_{0|0}} = {{\boldsymbol{x}}_0}$, ${{\boldsymbol{P}}_{0|0}} = {{\boldsymbol{P}}_0}$\; 
\textbf{Forward Pass:}\
\For{$t\leftarrow $1$ $ \KwTo $T$}{ 
Calculate ${{\boldsymbol{\tilde x}}_{t|t - 1}}$ and ${{\boldsymbol{P}}_{t|t-1}}$ using \eqref{prespttj1dqt}\;
Calculate ${{\boldsymbol{K}}_{t;MC}}$ and ${\phi _{\boldsymbol{R}}}$ using 
\begin{align}
\left\{ \begin{gathered}
  {{\boldsymbol{K}}_{t;MC}} = {\left( {{{\boldsymbol{P}}_{t|t - 1}} + {{\boldsymbol{H}}^{\text{T}}}{\phi _{\boldsymbol{R}}}{{\boldsymbol{R}}^{ - 1}}{\boldsymbol{H}}} \right)^{ - 1}}{{\boldsymbol{H}}^{\text{T}}}{\phi _{\boldsymbol{R}}}{{\boldsymbol{R}}^{ - 1}}, \hfill \\
  {\phi _{\boldsymbol{R}}}{\text{ = G}}\left( {{{\left\| {{{\boldsymbol{y}}_t} - {\boldsymbol{H}}{{\boldsymbol{x}}_{t|t - 1}}} \right\|}_{\boldsymbol{R}}}} \right); \hfill \\ 
\end{gathered}  \right.
\end{align}\

Calculate ${{\boldsymbol{\tilde x}}_{t|t}}$ and ${{\boldsymbol{P}}_{t|t}}$ using\
\begin{align}
\left\{ \begin{gathered}
  {{{\boldsymbol{\tilde x}}}_{t|t}} = {{{\boldsymbol{\tilde x}}}_{t|t - 1}}{\text{  +  }}{{\boldsymbol{K}}_{t;MC}}\left( {{{\boldsymbol{y}}_t} - {\boldsymbol{H}}{{{\boldsymbol{\tilde x}}}_{t|t - 1}}} \right), \hfill \\
  {{\boldsymbol{P}}_{t|t}} = {{\boldsymbol{K}}_{t;MC}}{{\boldsymbol{R}}_t}{\boldsymbol{K}}_{t;MC}^{\text{T}} +  \hfill \\
  \left( {{{\boldsymbol{I}}_n} - {{\boldsymbol{K}}_{t;MC}}{\boldsymbol{H}}} \right){{\boldsymbol{P}}_{t|t - 1}}{\left( {{{\boldsymbol{I}}_n} - {{\boldsymbol{K}}_{t;MC}}{\boldsymbol{H}}} \right)^{\text{T}}}; \hfill \\ 
\end{gathered}  \right.
\end{align}
}
\textbf{Backward Pass:}\
\For{$t\leftarrow $T - 1$ $ \KwTo $1$}{

Calculate ${\phi _{\boldsymbol{Q}}}$ and ${\boldsymbol{K}}_{t;MC}^b$ using
\begin{align}
\left\{ \begin{gathered}
  {\phi _{\boldsymbol{Q}}}{\text{ = G}}\left( {{{\left\| {{{{\boldsymbol{\tilde x}}}_{t + 1|T}} - {\boldsymbol{F}}{{{\boldsymbol{\tilde x}}}_{t|t}}} \right\|}_{\boldsymbol{Q}}}} \right), \hfill \\
  {\boldsymbol{K}}_{t;MC}^b = {\left( {{\boldsymbol{P}}_{t|t}^{ - 1} + {{\boldsymbol{F}}^{\text{T}}}{\phi _{\boldsymbol{Q}}}{{\boldsymbol{Q}}^{ - 1}}{\boldsymbol{F}}} \right)^{ - 1}}{{\boldsymbol{F}}^{\text{T}}}{\phi _{\boldsymbol{Q}}}{{\boldsymbol{Q}}^{ - 1}}; \hfill \\ 
\end{gathered}  \right.
\end{align}
Calculate ${{\boldsymbol{\tilde x}}_{t|T}}$ and ${{\boldsymbol{P}}_{t|T}}$ using\
\begin{align}
\left\{ \begin{gathered}
  {{{\boldsymbol{\tilde x}}}_{t|T}} = {{{\boldsymbol{\tilde x}}}_{t|t}} + {\boldsymbol{K}}_{t;MC}^b\left( {{{{\boldsymbol{\tilde x}}}_{t + 1|T}} - {\boldsymbol{F}}{{{\boldsymbol{\tilde x}}}_{t|t}}} \right), \hfill \\
  {{\boldsymbol{P}}_{t|T}} = {\boldsymbol{K}}_{t;MC}^b\left( {{{\boldsymbol{P}}_{t + 1|T}} - {{\boldsymbol{P}}_{t + 1|t}}} \right){\left( {{\boldsymbol{K}}_{t;MC}^b} \right)^{\text{T}}} \hfill \\
   + {{\boldsymbol{P}}_{t|t}}; \hfill \\ 
\end{gathered}  \right.
\end{align}
}
\end{algorithm}

\begin{table}
\centering
\caption{Computational complexities of the MEE-RTS algorithm.}\label{ccotmeerts}
\begin{tabular}{llll}
\hline
  &              &                                   & matrix inversion,\\ 
  & ${ \times }$ & ${ + {\text{ }}and{\text{ }} - }$ & Cholesky decomposition, \\
  &              &                                   & ,and exponentiation \\ \hline
\eqref{tr71qkj1axkjej}   & ${n^2}$              & ${n^2} - n$              & $0$       \\
\eqref{vkjxkCdYk}        & $2{n^3}$             & $2{n^3} - {n^2}$         & ${0}$     \\
\eqref{38xtTxmao}        & ${n^2}$              & ${n^2} + n$              & ${0}$     \\
\eqref{gainmeesmoother}  & $8{n^3}$             & $8{n^3} - 4{n^2}$        & $3O({n^3})$  \\
\eqref{ffAATT}      & $16{n^3}$            & $16{n^3} - 4{n^2}$       & ${0}$     \\
\eqref{FFigmatrix}    & $2n + 7$   & $2n + 1$        & $4$ \\
\eqref{Getigjetij}      & $2n + 7$   & $2n + 1$        & $4$  \\
\eqref{tKTtPtJ1IjN}      & $2{n^3}$             & $2{n^3}$                 & ${0}$     \\     \hline
\end{tabular}
\end{table}

\section{Simulation}\label{simulation}
In this section,  several numerical simulations are performed to demonstrate the performance of the proposed algorithms. In each run, 1000 samples are used to calculate the MSD of the MEE-RTS and MEE-ERTS smoothers. Simulation results are averaged over 300 independent Monte Carlo runs. The MSD \cite{HE2023108787} is defined as
\begin{equation}
\begin{split}  
MSD = 10{\log _{10}}\left( {{{\left\| {{{\boldsymbol{x}}_t} - {{{\boldsymbol{\tilde x}}}_{t|T}}} \right\|}^2}} \right).
\end{split}
\end{equation}

\subsection{Noise model}
A few non-Gaussian noise models are presented below before performing the simulation, such as mixed-Gaussian noise, $\alpha$-stable noise and Rayleigh noise.
\begin{itemize}
\item[1] Modeling mixed-Gaussian noise \cite{HE20221362} as follows:
\begin{align}\label{mixed-Gaussian}
r \sim \lambda \mathcal{N}\left( {{a_1},{\mu _1}} \right){\text{  +  }}\left( {1 - \lambda } \right)\mathcal{N}\left( {{a_2},{\mu _2}} \right),0 \leqslant \lambda  \leqslant 1,
\end{align}
where parameter $\lambda $ presents the mixing factor. The noise with a mixed-Gaussian distribution can be represented as $r \sim M\left( {\lambda ,{a_1},{a_2},{\mu _1},{\mu _2}} \right)$.
\item[2] The characteristic function of the ${\alpha}$-stable noise \cite{peng2017constrained} is defined as:
\begin{align}
\psi \left( t \right) = \exp \left\{ {j\vartheta t - \gamma {{\left| t \right|}^{{a_3}}}\left[ {1 + jb\operatorname{sgn} \left( t \right)S\left( {t,{a_3}} \right)} \right]} \right\},
\end{align}
with
\begin{align}
S\left( {t,{a_3}} \right) = \left\{ {\begin{array}{*{20}{l}}
  {\tan \left( {\frac{{{a_3}\pi }}{2}} \right),{\text{   }}{a_3} \ne {\text{1,}}} \\ 
  {\frac{2}{\pi }\log \left| t \right|,{\text{       }}{a_3}{\text{  =  1}}{\text{,}}} 
\end{array}} \right.
\end{align}
where ${{a_3}}$ is the characteristic factor, $b$ represents the symmetry parameter, ${\gamma}$ and $\vartheta $ are the dispersion parameter and location parameter. The noise with ${\alpha}$-stable distribution is expressed as $r \sim S\left( {{a_3},b,\gamma ,\vartheta } \right)$.
\item[3] The Rayleigh distribution can be represented by the probability density function shown below:
\begin{align}
r\left( t \right) = \frac{t}{{{\sigma ^2}}}\exp \left( { - \frac{{{t^2}}}{{2{\sigma ^2}}}} \right).
\end{align}
The noise with the Rayleigh distribution is written as ${v \sim R\left( \sigma  \right)}$.
\end{itemize}

\subsection{Example 1}
In this sub-section, a constant acceleration motion model \cite{CHEN2017} is considered. The state-space model are given by
\begin{align}
{{\boldsymbol{x}}_t} = \left[ {\begin{array}{*{20}{c}}
  1&{\Delta T}&{{{\Delta {T^2}} \mathord{\left/
 {\vphantom {{\Delta {T^2}} 2}} \right.
 \kern-\nulldelimiterspace} 2}} \\ 
  0&1&{\Delta T} \\ 
  0&0&1 
\end{array}} \right]{{\boldsymbol{x}}_{t - 1}} + {{\boldsymbol{q}}_{t - 1}},
\end{align}
and
\begin{align}
{{\boldsymbol{y}}_t} = \left[ {\begin{array}{*{20}{c}}
  {\begin{array}{*{20}{c}}
  1&0&0 
\end{array}} \\ 
  {\begin{array}{*{20}{c}}
  0&1&0 
\end{array}} 
\end{array}} \right]{{\boldsymbol{x}}_t} + {{\boldsymbol{r}}_t}.
\end{align}
Here, $\Delta T = 0.1sec$ represents sampling interval of the model, ${{\boldsymbol{x}}_t} = {\left[ {\begin{array}{*{20}{c}}
  {{x_{1;t}}}&{{x_{2;t}}}&{{x_{3;t}}} 
\end{array}} \right]^{\text{T}}}$ with ${{x_{1;t}}}$, ${{x_{2;t}}}$, and ${{x_{3;t}}}$ denote position, velocity, and acceleration respectively.

Five scenarios with different noise models are considered. The specific forms of process and measurement noises are shown below: 
\begin{align}
\left\{ {\begin{array}{*{20}{l}}
  {{q_{t;i}} \sim M\left( {0.9,0,0,0.01,25} \right),} \\ 
  {{r_{t;i}} \sim \mathcal{N}\left( {0,0.01} \right),} 
\end{array}} \right.
\end{align}
where ${{q_{t;i}}}$ and ${{r_{t;i}}}$ represent the $i$th element of ${{{\boldsymbol{q}}_t}}$ and ${{{\boldsymbol{r}}_t}}$. The second scenario is that process and measurement noises are both mixed-Gaussian noise, which takes the following form
\begin{align}
\left\{ \begin{gathered}
  {q_{t;i}} \sim M\left( {0.9,0,0,0.01,25} \right), \hfill \\
  {r_{t;i}} \sim M\left( {0.7,0,0,0.01,900} \right). \hfill \\ 
\end{gathered}  \right.
\end{align}
The third scenario is that both the process noise is mixed-Gaussian noise, and the measurement noise is a mixture of ${\alpha}$-stable noise and Gaussian noise with the following form
\begin{align}
\left\{ \begin{gathered}
  {q_{t;i}} \sim M\left( {0.9,0,0,0.01,25} \right), \hfill \\
  {r_{t;i}} \sim 0.9S\left( {1.25,1,0,0.5} \right){\text{ + 0}}{\text{.1}}\mathcal{N}\left( {0,900} \right). \hfill \\ 
\end{gathered}  \right.
\end{align}
where $0.9$ denotes the probability that the noise ${{r_{t;i}}}$ obeys the distribution of $S\left( {1.25,1,0,0.5} \right)$, and the representation below has a similar meaning.
The fourth simulation scenario has the following form:
\begin{align}
\left\{ {\begin{array}{*{20}{l}}
  {{q_{t;i}} \sim M\left( {0.9,0,0,0.01,25} \right),} \\ 
  {{r_{t;i}} \sim 0.7R\left( 2 \right) + 0.3\mathcal{N}\left( {0,900} \right).} 
\end{array}} \right.
\end{align}
The fifth scenario is that the process noise obeys a mixed-Gaussian distribution and the measurement noise is a multimodal distribution of the following form
\begin{align}
\left\{ {\begin{array}{*{20}{l}}
  {{q_{t;i}} \sim M\left( {0.95,0,0,0.01,25} \right),} \\ 
  {{r_{t;i}} \sim M\left( {0.6,2, - 2,0.01,100} \right).} 
\end{array}} \right.
\end{align}

The initial state ${{\boldsymbol{x}}_0}$, initial estimated value ${{\boldsymbol{\tilde x}}_{0|0}}$, and initial covariance  ${{\boldsymbol{P}}_{0|0}}$ of the above mentioned five scenarios are given by:
\begin{align}\label{x0bolapxx00}
\left\{ \begin{gathered}
  {{\boldsymbol{x}}_0} \sim \mathcal{N}\left( {0,{{\boldsymbol{I}}_n}} \right), \hfill \\
  {{{\boldsymbol{\tilde x}}}_{0|0}} \sim \mathcal{N}\left( {{{\boldsymbol{x}}_0},{\boldsymbol{Q}}} \right), \hfill \\
  {{\boldsymbol{P}}_{0|0}} = {{\boldsymbol{I}}_n}. \hfill \\ 
\end{gathered}  \right.
\end{align}

\begin{figure*}[!t]
\centering
\subfloat[]{\includegraphics[width=3.5in]{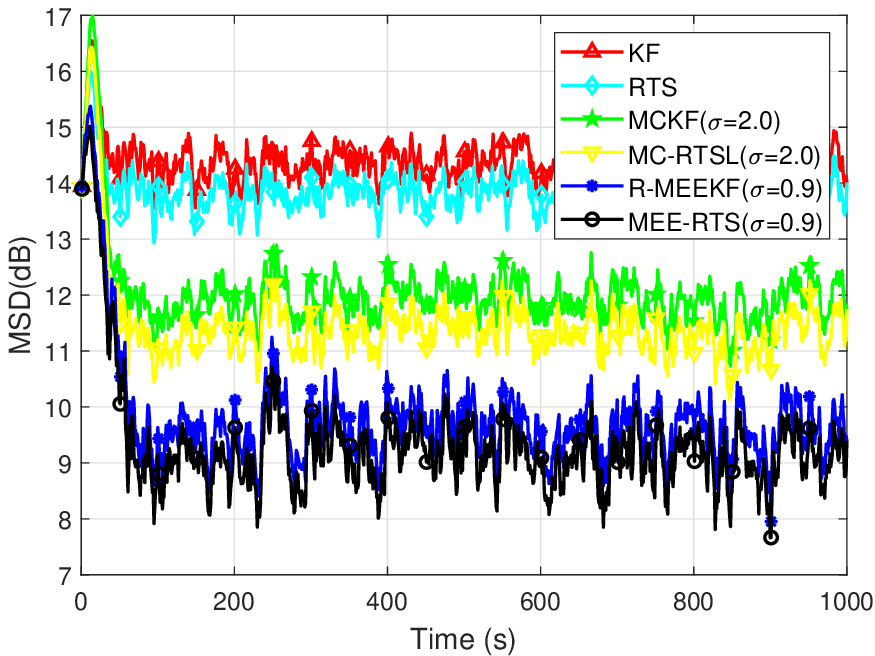}%
\label{Gaussianmixnoisepos}}
\hfil
\subfloat[]{\includegraphics[width=3.5in]{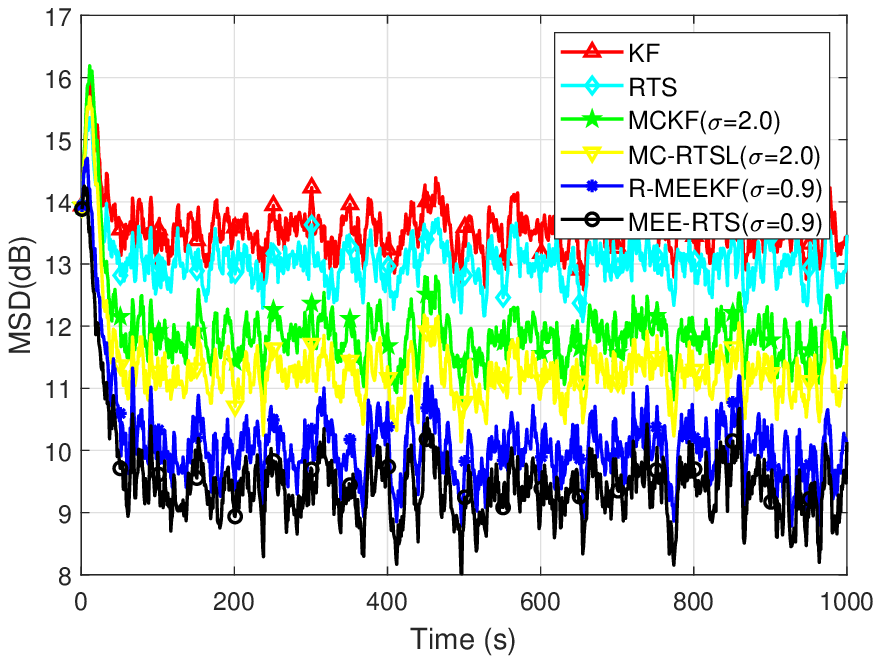}%
\label{Gaussianmixnoisevel}}
\hfil
\subfloat[]{\includegraphics[width=3.5in]{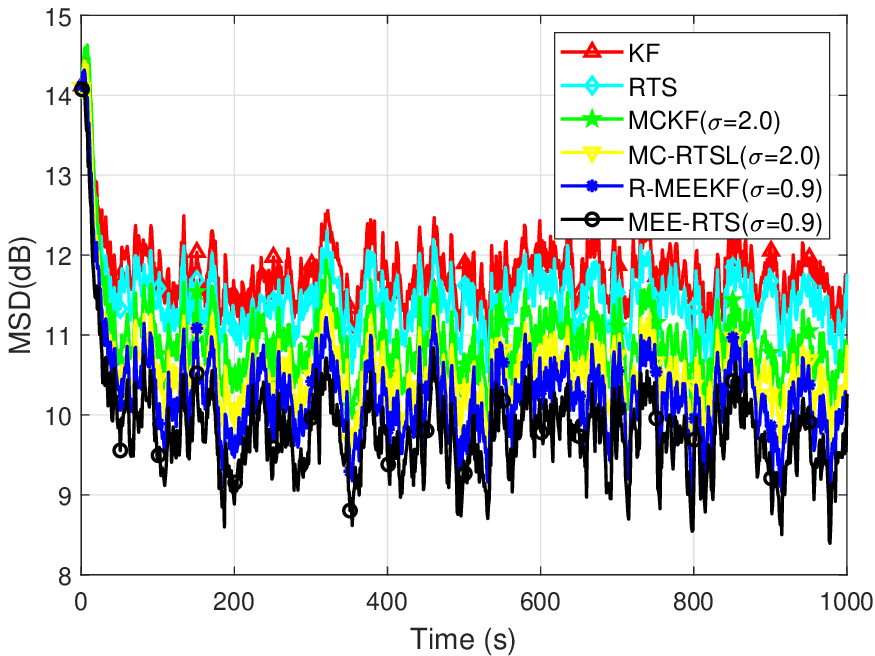}%
\label{Gaussianmixnoisea}}
\hfil
\subfloat[]{\includegraphics[width=3.5in]{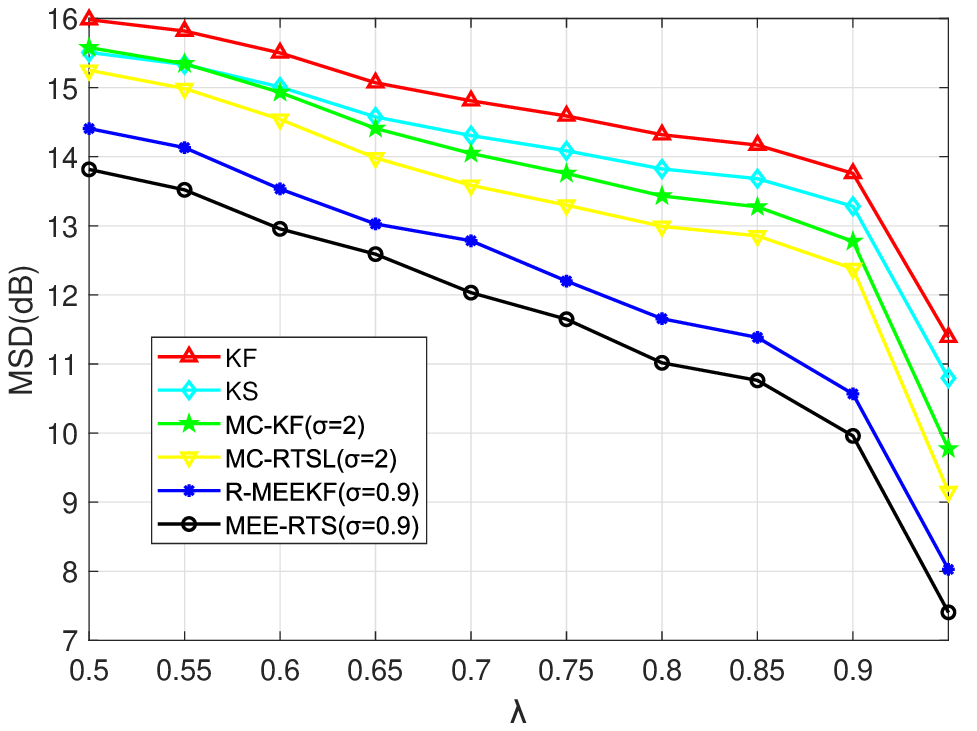}%
\label{Gaussianmixnoiseaxishuu}}
\caption{MSD of different algorithms in second scenario. (a) MSD of $x_1$. (b) MSD of $x_2$. (c) MSD of $x_3$. (d) The MSD with different $\lambda $.}
\label{two_mixture}
\end{figure*}

\begin{figure*}[!t]
\centering
\subfloat[]{\includegraphics[width=3.5in]{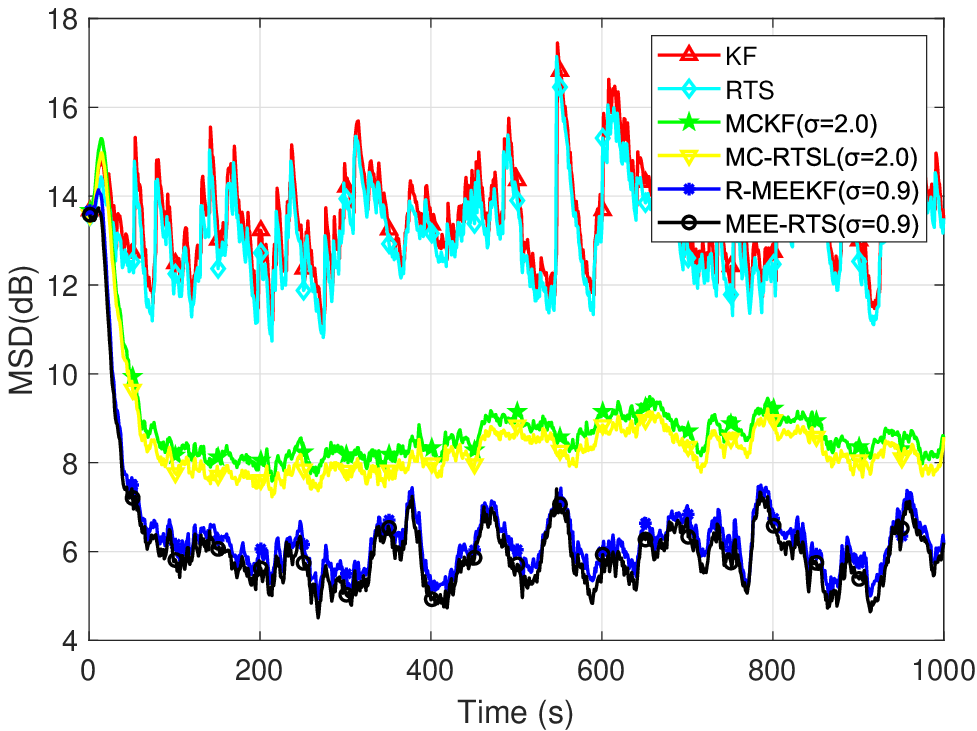}%
\label{figuremixalpha}}
\hfil
\subfloat[]{\includegraphics[width=3.5in]{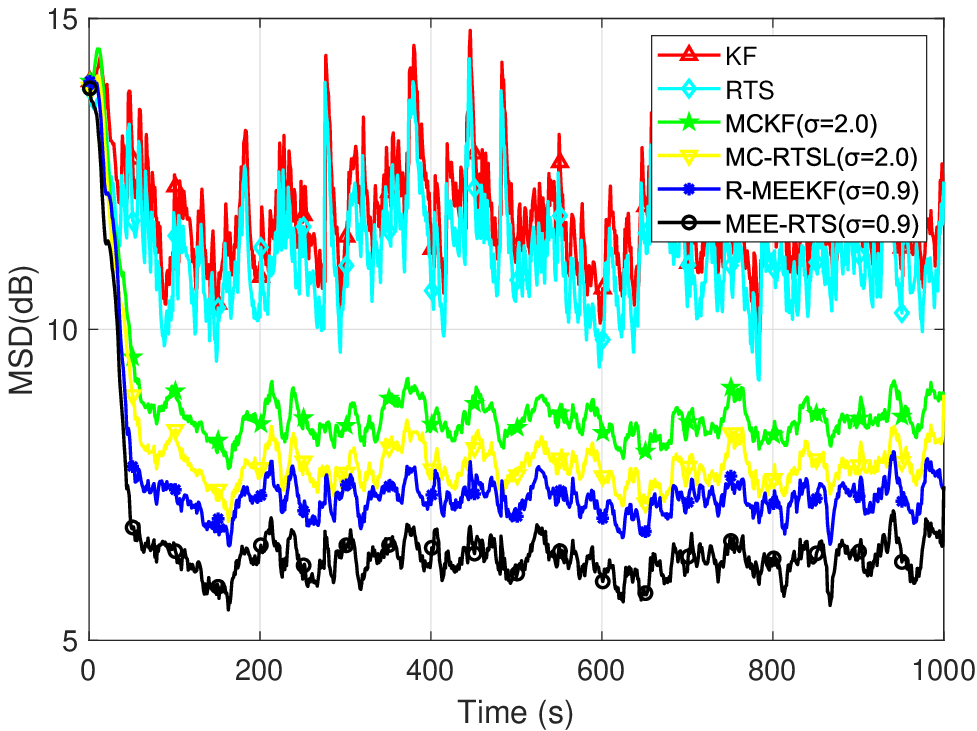}%
\label{alphanoisevel}}
\hfil
\subfloat[]{\includegraphics[width=3.5in]{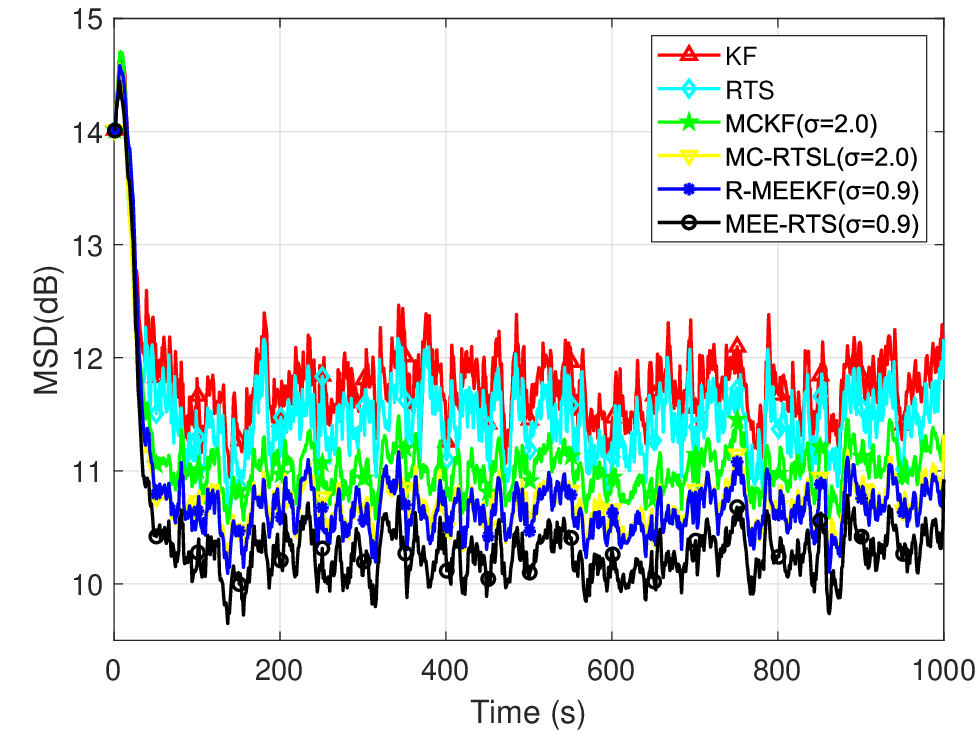}%
\label{alphanoisea}}
\hfil
\subfloat[]{\includegraphics[width=3.5in]{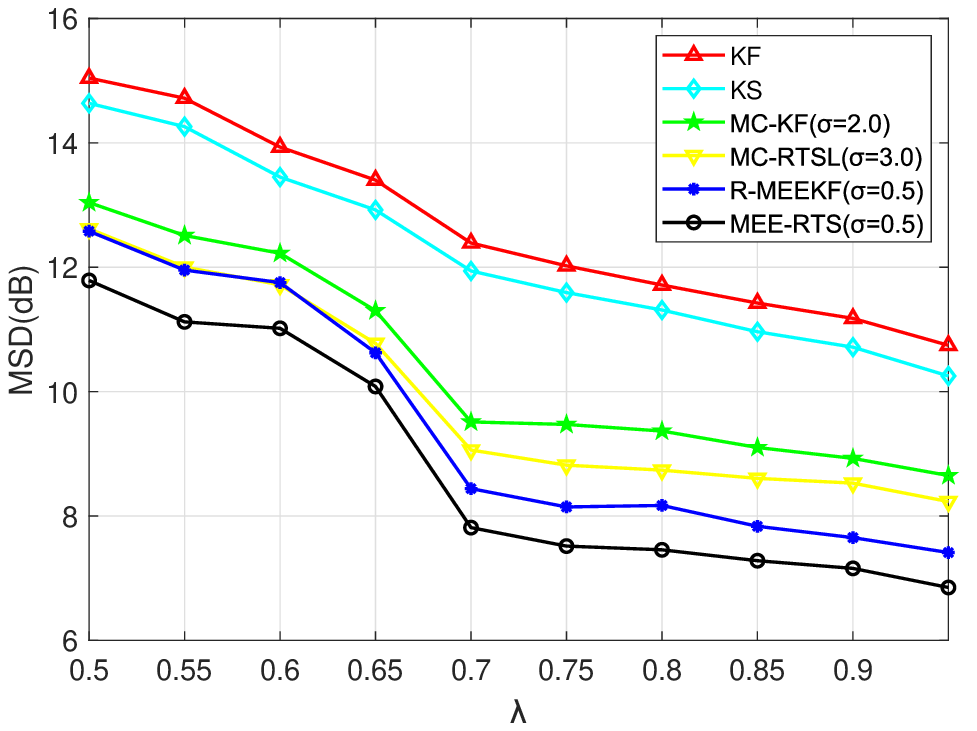}%
\label{alphanoisea1}}
\caption{The MSD of these considered algorithms in third scenario. (a) MSD of $x_1$. (b) MSD of $x_2$. (c) MSD of $x_3$. (d) The MSD with different $\lambda $.}
\label{mixalphamsd}
\end{figure*}

In these five scenarios, we compare the performance of the MEE-RTS smoother with that of the KF, RTS smoother \cite{rauch1965maximum}, MCKF \cite{CHEN2017}, MC-RTS smoother \cite{wang2020maximum}, R-MEEKF \cite{wang2021numerically} algorithms. These results are shown in Fig. \ref{two_mixture}, Fig. \ref{mixalphamsd}, and Table \ref{MSD_different_noise}.

First, the performance of the MEE-RTS smoother in a heavy-tailed noise environment is studied, and the simulation results are shown in Fig. \ref{two_mixture} and Fig. \ref{mixalphamsd}. The convergence curves of MSD of $x_1$, $x_2$, and $x_3$ are presented in Figs. \ref{Gaussianmixnoisepos}, \ref{Gaussianmixnoisevel}, and \ref{Gaussianmixnoisea}. Fig. \ref{Gaussianmixnoiseaxishuu} shows the variation curve of the MSD of ${\boldsymbol{x}}$ with the mixture coefficient $\lambda$ of measurement noise ${{r_t}}$. Some parameters of the considered algorithms are also presented in Fig. \ref{two_mixture}. It is obvious that the proposed MEE-RTS smoother performs best in the second scenario. From Fig. \ref{two_mixture}, one can witness that the performance of these algorithms degrades as the number of outliers increases. For the third scenario, the corresponding results and the parameter settings of these considered algorithms  are shown in Fig. \ref{mixalphamsd}, and the effect of the mixture coefficient $\lambda$ of the observed noise $r_i$ with respect to the performance of the MEE-RTS smoother was studied. Similar results to the second scenario can be obtained from Fig. \ref{mixalphamsd}. 

Then the robustness of the MEE-RTS smoother in different scenarios is investigated, and the simulation results are presented in Table \ref{MSD_different_noise}. The elements in Table \ref{MSD_different_noise} represent the MSD of $x_1$, $x_2$, and $x_3$ for the different algorithms in the considered scenarios, for example (-20.1, -20.1, -20.1) represents the MSD of $x_1$, $x_2$, and $x_3$ of the KF algorithm in the first scenario. In particular, $2.0$ in $0.9 (2.0)$ denotes the kernel width of the R-MEE and RTS algorithms in the fifth scenario. From Table \ref{MSD_different_noise}, we can get that 1): the MEE-RTS smoother performs best with the non-Gaussian noise; 2): in a Gaussian noise environment, the MEE-RTS algorithm performs nearly as well as the RTS algorithm. These two points show that the MEE-RTS smoother is robust in different scenarios.

\begin{table*}[htbp]
\centering
\caption{The steady-state MSD (dB) of different algorithms in different scenarios}\label{MSD_different_noise}
\resizebox{\textwidth}{!}
{
\begin{tabular}{lllllll}
\hline
Algorithm    & $\sigma$  & First scenario            & Second scenario        & Third scenario    & Fourth scenario    & Fifth scenario \\ \hline
KF           & N/A    & (-20.1,-20.1,-20.1)          & (11.5,14.8,12.0)       & (15.8,13.3,9.0)   & (13.3,14.8,14.4)   & (9.0,8.6,8.2)\\
RTS          & N/A    & (\textbf{-20.1,-20.2,-20.2}) & (14.4,11.4,10.9)       & (15.3,12.9,8.8)   & (14.3,13.8,12.9)   & (8.3,7.9,7.6)\\
MCKF         & 2.0    & (-19.2,-19.2,-19.2)          & (14.0,11.2,10.9)       & (12.5,11.7,8.6)   & (14.0,13.7,12.9)   & (8.8,8.6,7.9)\\
MC-RTS       & 2.0    & (-19.3,-19.3,-19.3)          & (13.5,10.3,10.1)       & (12.0,11.2,8.5)   & (13.4,13.1,12.4)   & (8.1,7.8,7.3)\\
R-MEEKF      & 0.9(2.0)    & (-19.0,-19.0,-19.0)          & (10.5,10.2,10.5)       & (9.0,9.4,8.8)     & (12.2,12.6,12.2)   & (8.5,8.5,7.7)\\
MEE-RTS      & 0.9(2.0)    & (-19.2,-19.1,-19.0)         & (\textbf{9.5,9.3,9.7}) & (\textbf{8.4,8.5,7.9}) & (\textbf{10.8,11.0,11.5}) & (\textbf{7.9,7.7,7.1})\\ \hline
\end{tabular}
}
\end{table*}

\begin{table*}[htbp]
\centering
\caption{The steady-state MSD (dB) of different algorithms with different kernel bandwidth}\label{MSD_kernel}
\resizebox{\textwidth}{!}
{
\begin{tabular}{lllllll}
\hline
Algorithm & $\sigma$ & First scenario            & Second scenario  & Third scenario            & Fourth scenario     & Fifth scenario\\   \hline
RTS       & N/A   & (\textbf{-20.1,-20.2,-20.2}) & (14.4,11.4,10.9) & (15.3,12.9,8.8)           & (14.3,13.1,12.4)    & (8.3,7.9,7.6)\\
MC-RTS    & 2     & (-19.3,-19.3,-19.3)          & (13.5,10.3,10.1) & (12.0,11.2,8.5)           & (13.4,13.1,12.4)    & (8.1,7.8,7.3)\\
MEE-RTS   & 0.1   & (-19.2,-19.1,-19.0)          & (17.1,14.6,13.3) & (18.3,17.3,15.4)           & (18.3,17.2,15.5)   & (13.7,13.2,12.1) \\
MEE-RTS   & 0.5   & (-19.2,-19.1,-19.0)          & (13.1,11.4,11.3) & (11.7,11.9,12.3)           & (11.9,11.7,12.1)   & (13.1,11.9,11.2)\\
MEE-RTS   & 0.9   & (-19.2,-19.1,-19.0           & (\textbf{9.5,9.3,9.7}) & (\textbf{8.4,8.5,7.9})   & (\textbf{10.8,11.0,11.5}) & (11.5,10.4,10.2)\\
MEE-RTS   & 2.0   & (-19.3,-19.3,-19.1)          & (13.3,9.5,9.2)   & (11.9,12.0,11.8)           & (12.0,11.9,11.7)   & (\textbf{7.9,7.7,7.1})\\
MEE-RTS   & 5.0   & (-19.4,-19.4,-19.3)          & (14.6,9.4,10.0)  & (13.9,13.6,12.8)           & (14.0,13.5,12.8)   & (8.9,8.7,8.5)\\
MEE-RTS   & 50.0  & (-19.5,-19.6,-19.7)          & (17.9,11.9,11.3) & (14.6,14.3,13.3)           & (14.7,14.3,13.4)   & (9.1,9.1,8.8)\\ 
MEE-RTS   & 100.0 & (-19.9,-19.9,-19.8)          & (18.3,13.5,12.1) & (15.2,15.1,15.1)           & (16.5,15.1,14.7)   & (9.2,9.5,9.8)\\ 
\hline
\end{tabular}
}
\end{table*}

\subsection{Example 2}
In this part, a vehicle tracking model \cite{9923771} is considered to verify the performance of the MEE-ETRS smoother, and the dataset \cite{citweb} of a vehicle is obtained from the the measurements of lidar and radar sensors. The state equation can be expressed as
\begin{align}
{{\boldsymbol{x}}_t} = \left[ {\begin{array}{*{20}{c}}
  1&0&{\Delta T}&0 \\ 
  0&1&0&{\Delta T} \\ 
  0&0&1&0 \\ 
  0&0&0&1 
\end{array}} \right]{{\boldsymbol{x}}_{t - 1}} + {{\boldsymbol{q}}_{t - 1}},
\end{align}
and the measurement equation of a radar is
\begin{align}
{{\boldsymbol{y}}_{t;radar}} = \left[ {\begin{array}{*{20}{c}}
  {\sqrt {x_{1;t}^2 + x_{2;t}^2} } \\ 
  {\arctan \left( {\frac{{{x_{2;t}}}}{{{x_{1;t}}}}} \right)} \\ 
  {\frac{{{x_{1;t}}{x_{3;t}} + {x_{2;t}}{x_{4;t}}}}{{\sqrt {x_{1;t}^2 + x_{2;t}^2} }}} 
\end{array}} \right] + {{\boldsymbol{r}}_{t;radar}},
\end{align}
and of a lidar is given by
\begin{align}
{{\boldsymbol{y}}_{t;lidar}} = \left[ {\begin{array}{*{20}{c}}
  {\begin{array}{*{20}{c}}
  1&0&0&0 
\end{array}} \\ 
  {\begin{array}{*{20}{c}}
  0&1&0&0 
\end{array}} 
\end{array}} \right]{{\boldsymbol{x}}_t} + {{\boldsymbol{r}}_{t;lidar}},
\end{align}
for lidar.
${{\boldsymbol{r}}_{t;radar}}$ and ${{\boldsymbol{r}}_{t;lidar}}$ are measurement noises of radar and lidar, and they obey distributions ${{\boldsymbol{r}}_{t;lidar}} \sim M\left( {\begin{array}{*{20}{c}}
  {0.9}&0&{0.01{{\boldsymbol{I}}_3}}&{9{{\boldsymbol{r}}_{radar}}} 
\end{array}} \right)$ and ${{\boldsymbol{r}}_{t;lidar}} \sim M\left( {\begin{array}{*{20}{c}}
  {0.9}&0&{0.01{{\boldsymbol{I}}_2}}&{9{{\boldsymbol{r}}_{lidar}}} 
\end{array}} \right)$. Here, the matrices ${{\boldsymbol{r}}_{radar}}$ and ${{\boldsymbol{r}}_{lidar}}$ are $diag\left[ {\begin{array}{*{20}{c}}
  1&{0.01}&{0.01} 
\end{array}} \right]$ and $diag\left[ {\begin{array}{*{20}{c}}
  1&{0.01} 
\end{array}} \right]$.

The state of vehicle can be represented as ${{\boldsymbol{x}}_t} = {\left[ {{x_{1;t}},{x_{2;t}},{x_{3;t}},{x_{4;t}}} \right]^{\text{T}}}$. ${{x_{1;t}}}$ and ${{x_{2;t}}}$ represent the position of $x$, $y$ axis, ${{x_{3;t}}}$ and ${{x_{4;t}}}$ denote the velocity of $x$, $y$ axis. The sampling interval is $\Delta T = 0.1\sec $, the covariance matrix of process noise is
\begin{align*}
{{\boldsymbol{Q}}} = \left[ {\begin{array}{*{20}{c}}
  {\tfrac{{\Delta {T^2}}}{4}}&0&{\tfrac{{\Delta {T^3}}}{2}}&0 \\ 
  0&{\tfrac{{\Delta {T^2}}}{4}}&0&{\tfrac{{\Delta {T^3}}}{2}} \\ 
  {\tfrac{{\Delta {T^3}}}{2}}&0&{\Delta {T^2}}&0 \\ 
  0&{\tfrac{{\Delta {T^3}}}{2}}&0&{\Delta {T^2}} 
 \end{array}} \right].
\end{align*}
The initial values of ${{\boldsymbol{x}}_0}$, ${{\boldsymbol{\tilde x}}_{0|0}}$, and ${{\boldsymbol{P}}_{0|0}}$ are the same as \eqref{x0bolapxx00}. The covariance matrix of measurement noise is set to ${{\boldsymbol{R}}_t} = \left[ {0.09,0.05,0.09} \right]$.
 
\begin{figure}
\centerline{\includegraphics[width=0.5\textwidth]{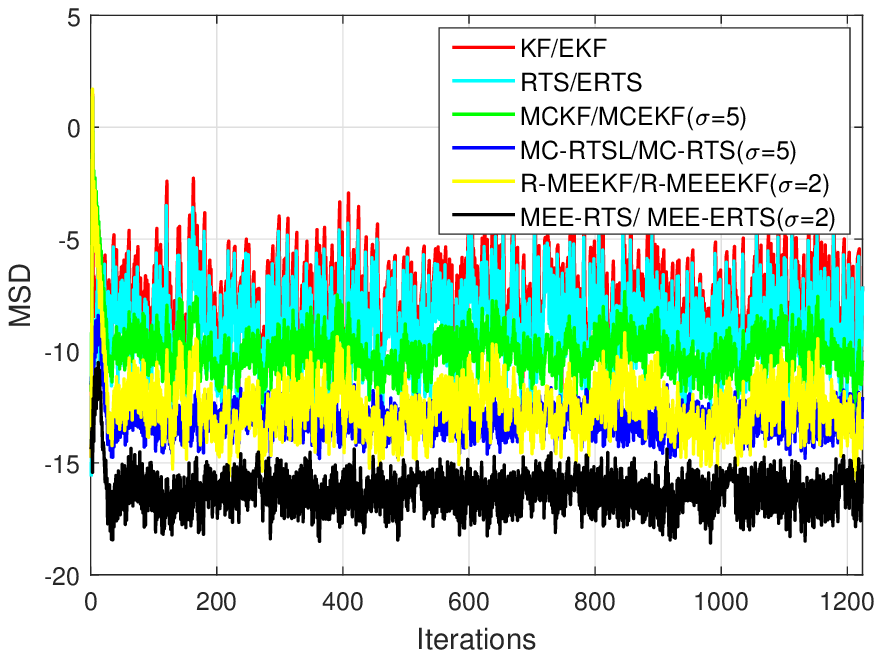}}
\caption{MSD of different algorithms.}\label{tracking}
\end{figure}

The variation curves of the MSD of different algorithms are presented in Fig. \ref{tracking}. According to Fig. \ref{tracking}, one can obtain that the MEE-ERTS smoother outperforms several algorithms in terms of steady-state error.

\subsection{Parameter selection} \label{para_perfro_mee}
In this part, the influence of the critical parameters $\sigma$ and $\tau$ on the performance of the MEE-RTS smoother is investigated, and guidance is given on the choice of parameters.

\begin{figure*}[!t]
\centering
\subfloat[]{\includegraphics[width=3.5in]{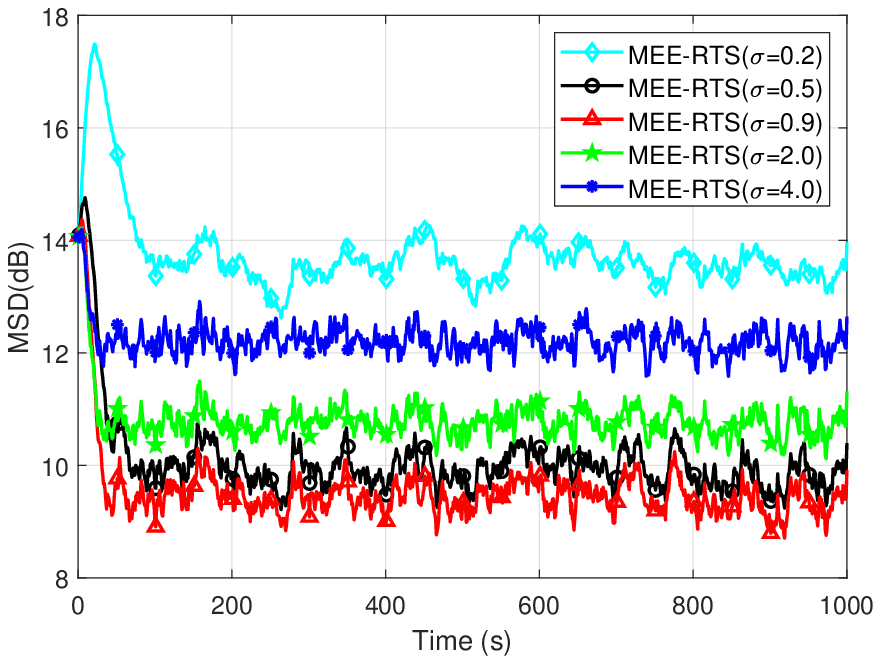}%
\label{MEERTSsigma}}
\hfil
\subfloat[]{\includegraphics[width=3.5in]{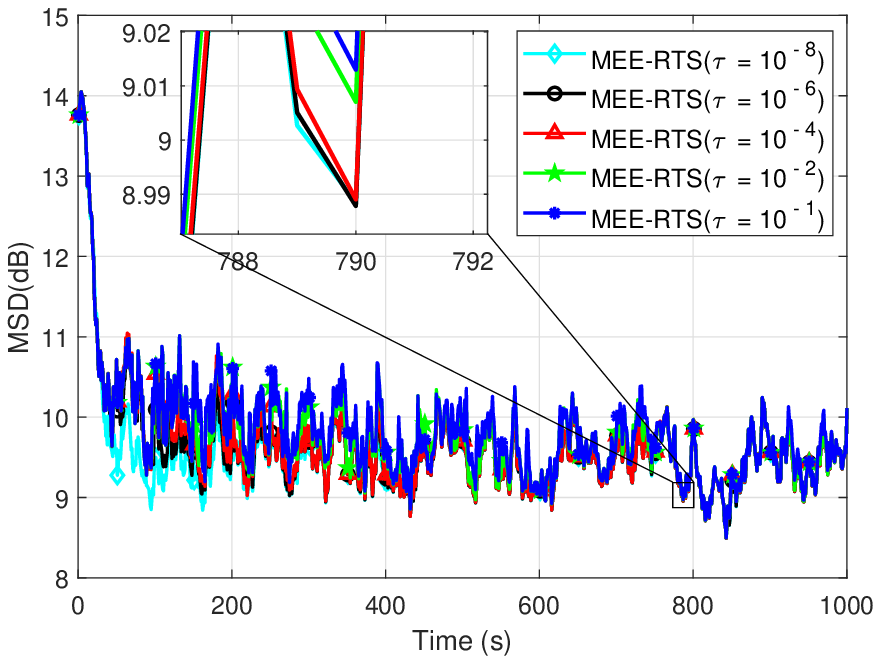}%
\label{meertstau}}
\caption{MSD of MEE-RTS smoother with different $\sigma$. (a) MSD of MEE-RTS smoother with different $\sigma$. (b) MSD of the MEE-RTS smoother with different $\tau $.}
\label{MEERTSTau}
\end{figure*}

The influence of the kernel bandwidth $\sigma$ on the performance of the MEE-RTS smoother is investigated in the above five scenarios, and the simulation results are displayed in Table \ref{MSD_kernel} and Fig. \ref{MEERTSsigma}. As seen in Table \ref{MSD_kernel}, one can be inferred that the performance of the MEE-RTS smoother first increases and then decreases in the increase of the parameter $\sigma$ with the presence of non-Gaussian noise. The MEE-RTS smoother can yield the required performance when $\sigma$ is close to 0.9 or 2.0. In a Gaussian noise environment, the performance of the MEE-RTS smoother improves with the increasing $sigma$ but does not exceed that of the RTS smoother and KF algorithms. The relationship between algorithm performance and kernel bandwidth guides the selection of kernel bandwidth. The selection scheme about $\sigma$ is as follows: 1) a larger $\sigma$ should be set in a Gaussian noise environment; 2) the optimal kernel width is different in different heavy-tailed noise scenarios, and suitable $\sigma$ is 0.9 or 2.0 in the considered scenarios.

The influence of threshold $\tau$ with respect to the performance of the MEE-RTS algorithm is also investigated in the presence of non-Gaussian noise. The simulation results are presented in Fig. \ref{meertstau} and Table \ref{MSD_tau} with different $\tau$. In the second scenario, Fig. \ref{meertstau} presents the MSD of MEE-RTS smoother with different $\tau$ and $\sigma = 0.9$. The MSD, computational time, and the number $M$ of the FPI method of RTS, MC-RTSL, and MEE-RTS smoothers are presented in Table \ref{MSD_tau}. These smoothers are measured using MATLAB 2020a, which works on an i5-8400 and a 2.80GHz CPU. From Fig. \ref{MEERTSTau} and Table \ref{MSD_tau}, the performance of the MEE-RTS smoother in steady-state MSD decreases to some extent as the threshold decreases in non-Gaussian noise environments, however, the cost (the computational complexity and computational time) of such a small increase in performance is expensive. Hence, the choice of threshold should trade off the steady-state MSD and cost, with smaller thresholds being chosen to improve the performance of the algorithm where real-time requirements are low, and larger thresholds or ARM being used in the opposite direction.

\begin{table*}
\centering
\caption{The steady-state MSD, computational time, and the number of the FPI method of different algorithms with different $\tau$}\label{MSD_tau}
\resizebox{\textwidth}{!}
{
\begin{tabular}{lllllllllll}
\hline
          & $\tau$            & \multicolumn{3}{c}{Second scenario} & \multicolumn{3}{c}{Third scenario} & \multicolumn{3}{c}{Fourth scenario}  \\ \hline
Algorithm &                   & MSD   & time (sec)                                       & M        & MSD     & time (sec)                                      & M        & MSD     & time (sec)                                      & M             \\
RTS       & N/A               & 13.29 & ${\text{1}}{\text{.685}} \times {10^{ - 5}}$     & N/A      & 12.94   & ${\text{1}}{\text{.643}} \times {10^{ - 5}}$   & N/A      & 13.52    & ${\text{1}}{\text{.691}} \times {10^{ - 5}}$    & N/A           \\
MC-RTSL   & N/A               & 11.07 & ${\text{7}}{\text{.499}} \times {10^{ - 5}}$     & N/A      & 10.89   & ${\text{1}}{\text{.679}} \times {10^{ - 5}}$   & N/A      & 12.81    & ${\text{1}}{\text{.656}} \times {10^{ - 5}}$    & N/A           \\
MEE-RTS   & ${10^{ - 1}}$     & 9.52  & ${\text{2}}{\text{.322}} \times {10^{ - 4}}$     & 1.0249   & 8.99    & ${\text{2}}{\text{.302}} \times {10^{ - 4}}$   & 1.0225   & 10.93    & ${\text{2}}{\text{.337}} \times {10^{ - 4}}$    & 1.0101        \\
MEE-RTS   & ${10^{ - 2}}$     & 9.51  & ${\text{2}}{\text{.562}} \times {10^{ - 4}}$     & 1.2128   & 8.94    & ${\text{2}}{\text{.563}} \times {10^{ - 4}}$   & 1.1962   & 10.91    & ${\text{2}}{\text{.530}} \times {10^{ - 4}}$    & 1.1893        \\
MEE-RTS   & ${10^{ - 3}}$     & 9.49  & ${\text{3}}{\text{.155}} \times {10^{ - 4}}$     & 1.6599   & 8.93    & ${\text{3}}{\text{.130}} \times {10^{ - 4}}$   & 1.6423   & 10.89    & ${\text{3}}{\text{.104}} \times {10^{ - 4}}$    & 1.6098        \\
MEE-RTS   & ${10^{ - 4}}$     & 9.48  & ${\text{3}}{\text{.419}} \times {10^{ - 4}}$     & 1.9380   & 8.91    & ${\text{3}}{\text{.349}} \times {10^{ - 4}}$   & 1.9413   & 10.88    & ${\text{3}}{\text{.324}} \times {10^{ - 4}}$    & 1.8310        \\
MEE-RTS   & ${10^{ - 5}}$     & 9.47  & ${\text{3}}{\text{.849}} \times {10^{ - 4}}$     & 2.3892   & 8.90    & ${\text{3}}{\text{.530}} \times {10^{ - 4}}$   & 2.4786   & 10.87    & ${\text{3}}{\text{.687}} \times {10^{ - 4}}$    & 2.2041        \\
MEE-RTS   & ${10^{ - 6}}$     & 9.46  & ${\text{4}}{\text{.453}} \times {10^{ - 4}}$     & 2.9323   & 8.89    & ${\text{4}}{\text{.543}} \times {10^{ - 4}}$   & 3.0513   & 10.87    & ${\text{4}}{\text{.155}} \times {10^{ - 4}}$    & 2.6500        \\
MEE-RTS   & ${10^{ - 7}}$     & 9.45  & ${\text{4}}{\text{.931}} \times {10^{ - 4}}$     & 3.4424   & 8.88    & ${\text{5}}{\text{.177}} \times {10^{ - 4}}$   & 3.6535   & 10.86    & ${\text{7}}{\text{.701}} \times {10^{ - 4}}$    & 3.1446        \\
MEE-RTS   & ${10^{ - 8}}$     & 9.45  & ${\text{5}}{\text{.468}} \times {10^{ - 4}}$     & 3.9648   & 8.88    & ${\text{5}}{\text{.740}} \times {10^{ - 4}}$   & 4.2352   & 10.86    & ${\text{5}}{\text{.076}} \times {10^{ - 4}}$    & 3.4870        \\ \hline
\end{tabular}
}
\end{table*}

\section{Conclusion}\label{conclusion}
In this study, two new RTS-type smoothers based on the MEE criterion were developed for estimating the state of the linear and nonlinear systems with non-Gaussian noise. Some theoretical analysis of the MEE-RTS smoother were given, and the results of these analyses have indicated that the MEE-RTS smoother is stable. Numerical simulation results have indicated that the proposed smoothers perform better than several existing smoothers with non-Gaussian noise.

The choice of threshold $\tau$ should trade off the steady-state MSD and computational complexity of the MEE-RTS smoother, with smaller threshold being chosen to improve the performance of the algorithm where real-time requirements are low, and larger thresholds or ARM being used in the opposite direction. Furthermore, the kernel width $\sigma$ has a significant influence on the steady-state MSD. Kernel width that are either too large or too small can degrade the performance of the MEE-RTS smoother and should be adjusted empirically. 

The MEE-ERTS smoother is also developed for nonlinear systems. However, retaining only the first-order Taylor expansion term inevitably introduces errors due to linearization. To extend the application, there is a need to develop an RTS-type smoother based on the MEE criterion with higher accuracy for estimating the state of nonlinear systems, which is our future work.

\section{Acknowledgments}\label{acknowledgments}
This work was funded by the NNSFC with Nos. 51975107 and 62103083, Sichuan Science and Technology Major Project No.2019ZDZX0020, and Sichuan Science and Technology Program, No. 2022YFG0343. The first two authors contributed equally to this study.

\appendix
\section{Derivation of the formula \eqref{tKTtPtJ1IjN}} \label{deadapin} 
The covariance of the errors ${{\boldsymbol{\xi }}_{s;t}}$ and ${{\boldsymbol{\xi }}_{f;t}}$ is defined as 
\begin{align}
\left\{ \begin{gathered}
  {\text{E}}\left[ {{{\boldsymbol{\xi }}_{s;t}}{\boldsymbol{\xi }}_{s;t}^{\text{T}}} \right] = Cov\left( {{{\boldsymbol{\xi }}_{s;t}}} \right) = {{\boldsymbol{P}}_{t|T}}, \hfill \\
  {\text{E}}\left[ {{{\boldsymbol{\xi }}_{f;t}}{\boldsymbol{\xi }}_{f;t}^{\text{T}}} \right] = Cov\left( {{{\boldsymbol{\xi }}_{f;t}}} \right) = {{\boldsymbol{P}}_{t|t}}. \hfill \\ 
\end{gathered}  \right.
\end{align} Combining ${{\boldsymbol{\xi }}_{s;t}} = {{\boldsymbol{x}}_t} - {{\boldsymbol{\tilde x}}_{t|T}}$ and ${{\boldsymbol{\xi }}_{f;t}} = {{\boldsymbol{x}}_t} - {{\boldsymbol{\tilde x}}_{t|t}}$, \eqref{38xtTxmao} can be rewritten as
\begin{align}\label{xovApoint1}
{{\boldsymbol{\xi }}_{f;t}} + {\boldsymbol{K}}_t^b{{\boldsymbol{\tilde x}}_{t + 1|t}} = {{\boldsymbol{\xi }}_{s;t}} + {\boldsymbol{K}}_t^b{{\boldsymbol{\tilde x}}_{t + 1|T}}.
\end{align}
Taking covariance on both sides of \eqref{xovApoint1}, one can obtain
\begin{align}\label{wumingp}
\begin{gathered}
  {\text{E}}\left[ \begin{gathered}
  {{\boldsymbol{\xi }}_{f;t}}{\boldsymbol{\xi }}_{f;t}^{\text{T}} + {{\boldsymbol{\xi }}_{f;t}}{\boldsymbol{\tilde x}}_{t + 1|t}^{\text{T}}{\left( {{\boldsymbol{K}}_t^b} \right)^{\text{T}}} +  \hfill \\
  {\boldsymbol{K}}_t^b{{{\boldsymbol{\tilde x}}}_{t + 1|t}}{\boldsymbol{\xi }}_{f;t}^{\text{T}} + {\boldsymbol{K}}_t^b{{{\boldsymbol{\tilde x}}}_{t + 1|t}}{\boldsymbol{\tilde x}}_{t + 1|t}^{\text{T}}{\left( {{\boldsymbol{K}}_t^b} \right)^{\text{T}}} \hfill \\ 
\end{gathered}  \right] =  \hfill \\
  {\text{E}}\left[ \begin{gathered}
  {{\boldsymbol{\xi }}_{s;t}}{\boldsymbol{\xi }}_{s;t}^{\text{T}} + {{\boldsymbol{\xi }}_{s;t}}{\boldsymbol{\tilde x}}_{t + 1|T}^{\text{T}}{\left( {{\boldsymbol{K}}_t^b} \right)^{\text{T}}} +  \hfill \\
  {\boldsymbol{K}}_t^b{{{\boldsymbol{\tilde x}}}_{t + 1|T}}{\boldsymbol{\xi }}_{s;t}^{\text{T}} + {\boldsymbol{K}}_t^b{{{\boldsymbol{\tilde x}}}_{t + 1|T}}{\boldsymbol{\tilde x}}_{t + 1|T}^{\text{T}}{\left( {{\boldsymbol{K}}_t^b} \right)^{\text{T}}} \hfill \\ 
\end{gathered}  \right]. \hfill \\ 
\end{gathered} 
\end{align}
Since ${\text{E}}\left[ {{{\boldsymbol{\xi }}_{f;t}}{\boldsymbol{\tilde x}}_{t + 1|t}^{\text{T}}} \right] = 0$ and ${\text{E}}\left[ {{{\boldsymbol{\xi }}_{s;t}}{\boldsymbol{\tilde x}}_{t + 1|T}^{\text{T}}} \right] = 0$ proposed in \cite{rauch1965maximum}, \eqref{wumingp} can be simplified to
\begin{align}\label{kbtTxtj1TT}
\begin{gathered}
  {\text{E}}\left[ {{{\boldsymbol{\xi }}_{f;t}}{\boldsymbol{\xi }}_{f;t}^{\text{T}}} \right] + {\boldsymbol{K}}_t^b{\boldsymbol{F}}{\text{E}}\left[ {{{{\boldsymbol{\tilde x}}}_{t|t}}{\boldsymbol{\tilde x}}_{t|t}^{\text{T}}} \right]{{\boldsymbol{F}}^{\text{T}}}{\left( {{\boldsymbol{K}}_t^b} \right)^{\text{T}}} \hfill \\
   = {\text{E}}\left[ {{{\boldsymbol{\xi }}_{s;t}}{\boldsymbol{\xi }}_{s;t}^{\text{T}}} \right] + {\text{E}}\left[ {{\boldsymbol{K}}_t^b{{{\boldsymbol{\tilde x}}}_{t + 1|T}}{\boldsymbol{\tilde x}}_{t + 1|T}^{\text{T}}{{\left( {{\boldsymbol{K}}_t^b} \right)}^{\text{T}}}} \right]. \hfill \\ 
\end{gathered} 
\end{align}
Substituting the fact $Cov\left( {{{{\boldsymbol{\tilde x}}}_{t|t}}} \right) = Cov\left( {{{\boldsymbol{x}}_t}} \right) - {{\boldsymbol{P}}_{t|t}}$ and $Cov\left( {{{{\boldsymbol{\tilde x}}}_{t + 1|T}}} \right) = Cov\left( {{{\boldsymbol{x}}_{t + 1}}} \right) - {{\boldsymbol{P}}_{t + 1|T}}$ into \eqref{kbtTxtj1TT}, and we can obtain \eqref{tKTtPtJ1IjN}.

\bibliographystyle{IEEEtran}
\bibliography{ref}

\vfill

\end{document}